\documentclass[11pt]{article}
\usepackage{jheppub}

\usepackage{amssymb}
\usepackage{amsmath}
\usepackage{amsfonts}

\usepackage{graphicx, subfigure, placeins, float}

\usepackage{array} 
\usepackage{longtable} 
\usepackage{colortab} 
\usepackage{colortbl}
\usepackage{arydshln}

\usepackage{dsfont,bbm}
\usepackage{mathrsfs}  
\usepackage{xfrac}
\usepackage{slashed}
\usepackage{mathabx}
\usepackage{enumitem}
\usepackage{shuffle}
\usepackage{stmaryrd}

\makeatletter
\newcommand\xleftrightarrow[2][]{%
  \ext@arrow 9999{\longleftrightarrowfill@}{#1}{#2}}
\newcommand\longleftrightarrowfill@{%
  \arrowfill@\leftarrow\relbar\rightarrow}
\makeatother

\newcommand{\Rome}[1]{\uppercase\expandafter{\romannumeral#1}}

\newcommand{\itbf}[1]{\textbf{\textit{#1}}}

\newcolumntype{C}[1]{>{\centering\arraybackslash}m{#1}}

\makeatletter
\newcommand{\mathleft}{\@fleqntrue\@mathmargin0pt}
\newcommand{\mathcenter}{\@fleqnfalse}
\makeatother

\newcommand{\pl}{\text{PL}}
\newcommand{\np}{\text{NP}}

\usepackage{extarrows}

\allowdisplaybreaks[4]

\title{A high-precision result for a full-color three-loop three-point form factor in ${\cal N}=4$ SYM}

\author[a]{Xin Guan,}
\emailAdd{guanxin0507@pku.edu.cn}
\author[b]{Guanda Lin,}
\emailAdd{linguandak@pku.edu.cn}
\author[c]{Xiao Liu,}
\emailAdd{xiao.liu@physics.ox.ac.uk}
\author[a,d]{Yan-Qing Ma,}
\emailAdd{yqma@pku.edu.cn}
\author[e,f,g]{and Gang Yang}
\emailAdd{yangg@itp.ac.cn}
\affiliation[a]{School of Physics, Peking University, Beijing 100871, China}
\affiliation[b]{Department of Physics, University of California, Berkeley, CA 94720, U.S.A.}
\affiliation[c]{Rudolf Peierls Centre for Theoretical Physics, Clarendon Laboratory, Parks Road, Oxford OX1 3PU, UK}
\affiliation[d]{Center for High Energy Physics, Peking University, Beijing 100871, China}
\affiliation[e]{CAS Key Laboratory of Theoretical Physics, Institute of Theoretical Physics, \\Chinese Academy of Sciences, Beijing 100190, China}
\affiliation[f]{School of Fundamental Physics and Mathematical Sciences, Hangzhou Institute for Advanced Study, UCAS, Hangzhou 310024, China}
\affiliation[g]{International Centre for Theoretical Physics Asia-Pacific, Beijing/Hangzhou, China}

\abstract{
We perform a high-precision computation of the three-loop three-point form factor of the stress-tensor supermultiplet in ${\cal N}=4$ SYM.
Both the leading-color and sub-leading-color form factors are expanded in terms of simple integrals. 
We compute the complete set of integrals at a special kinematic point with very high precision using  {\tt AMFlow}.
The high-precision leading-color result enables us to obtain the analytic form of a numerical constant in the three-loop BDS ansatz, which is previously known only numerically. The high-precision values of the non-leading-color finite remainder
as well as all integrals are also presented, which can be valuable for future use.
}

\begin{document}

\maketitle

\setcounter{footnote}{0}

\section{Introduction}

Significant advances has been made in perturbative high loop computations in quantum field theory in recent years. 
As one such example, in this paper, we present a non-trivial high-precision computation of the full-color three-loop three-point form factor of the stress-tensor supermultiplet in ${\cal N}=4$ SYM.

The complete integrand of the three-loop form factor has been previously derived in \cite{Lin:2021kht, Lin:2021qol}, utilizing the color-kinematics duality \cite{Bern:2008qj, Bern:2010ue} and the unitary-cut method \cite{Bern:1994zx, Bern:1994cg, Britto:2004nc}.
The evaluation of three-loop form factor requires considering all possible planar and non-planar topologies. 
While there has been progress in analytically evaluating three-loop master integrals \cite{DiVita:2014pza, Canko:2021xmn, Henn:2023vbd}, the complete set of integrals for the problem at hand is still unavailable.
A previous attempt to evaluate these integrals using FIESTA \cite{Smirnov:2015mct} and pySecDec \cite{Borowka:2017idc} was made in \cite{Lin:2021qol}, where it required substantial computational resources yet yielded results of relatively low precision. 

The  {\tt AMFlow} method \cite{Liu:2017jxz, Liu:2021wks, Liu:2022mfb, Liu:2022chg} offers a viable solution to compute all the three-loop integrals with exceptionally high precision. 
In this paper we present the explicit results obtained using  {\tt AMFlow} at a special kinematic point. 
As we will see,  {\tt AMFlow} not only dramatically enhances the computational efficiency but also gives significantly more accurate results.
It is also worth mentioning that, while we focus on a single kinematic point in this study, the generalization to other kinematic points is straightforward.

Having the integral results, we obtain the full form factor which exhibits the correct infrared divergence. For the planar form factor, our high-precision result, together with the three-loop remainder \cite{Dixon:2020bbt}, enables us to determine the analytic expression for a constant appearing in the three-loop Bern-Dixon-Smirnov (BDS) ansatz \cite{Bern:2005iz}, a quantity previously known only in numerical terms \cite{Spradlin:2008uu}. Additionally, we present, for the first time, the high-precision value of the non-planar finite remainder.

The rest of the paper is organized as follows. In Section~\ref{sec:FF}, we set up the convention and present the integral expansion of the three-loop form factor. Section~\ref{sec:results} discusses the integration and the full form factor results. By subtracting the infrared divergences, we obtain the planar and non-planar finite remainders with high-precision values and determine the analytic expression of a coefficient in the BDS ansatz. 
A summary and outlook are provided in Section~\ref{sec:summary}.
In Appendix~\ref{ap:integrals}, we give the definition of all the integrals as well the as the integrated results.
For the reader's convenience, all results are also provided in the ancillary file.

\section{Form factor integrand}\label{sec:FF}

In this section, we first define the form factor and set up the convention. Then we give the three-loop form factor using a basis integral expansion. 

The quantity we consider is the three-point form factor of the stress tensor supermultiplet in ${\cal N}=4$ SYM:
\begin{equation}
    \itbf{F}_{\mathcal{T},3} = \int d^{D} x e^{-i q \cdot x}\langle \Phi_1^{a_1}(p_1) \Phi_2^{a_2}(p_2) \Phi_3^{a_3}(p_3)|\mathcal{T}(x)| 0\rangle \,,
\end{equation}
where $\Phi_i$'s represent on-shell superfields with $p_i^2=0, i = 1,2,3,$ and $q^2 = (p_1+p_2+p_3)^2 \neq0$. The operator $\mathcal{T}$ is the chiral stress-tensor supermultiplet, see \cite{Eden:2011yp,Brandhuber:2011tv}. In practice and without loss of generality, one can simply consider the bosonic component form factor with half-BPS operator ${\rm tr}(\phi_{12}^2)$ and the external states as 
\begin{equation}
    \itbf{F}_{\operatorname{tr}(\phi^2),3} = \int d^{D} x e^{-i q \cdot x} \langle \phi_{12}^{a_1}(p_1) \phi_{12}^{a_2}(p_2) g_+^{a_3}(p_3)|{\rm tr}(\phi_{12}^2)| 0\rangle \,.
\end{equation}

The perturbative loop expansion of form factor is 
\begin{equation}
\itbf{F}_{\operatorname{tr}(\phi^2),3} = \itbf{F}^{\rm tree}_{\operatorname{tr}(\phi^2),3} \, \sum_{\ell} \, g^{2\ell}  \itbf{I}^{(\ell)}_{\operatorname{tr}(\phi^2),3} \,,
\end{equation}
in which $\ell$ is the number of loops, the tree-level form factor is
\begin{equation}
\itbf{F}^{\rm tree}_{\operatorname{tr}(\phi^2),3} =  f^{a_1 a_2 a_3} {\langle 1 2\rangle^2 \over \langle 12 \rangle \langle 23 \rangle \langle 31 \rangle}  \,,
\end{equation}
and we introduce a modified 't Hooft coupling 
\begin{equation}
g^2 = {g_{{\cal N}=4}^2 N_c  \over 16\pi^2} (4\pi e^{-\gamma_E})^\epsilon \,,
\end{equation}
where $g_{{\cal N}=4}$ is the bare coupling of the model, $N_c$ is the number of colors, $\gamma_E$ is Euler’s constant, and $\epsilon = (4-D)/2$ is the parameter of dimensional regularization.

The full-color three-loop correction was obtained in \cite{Lin:2021qol} which can be organized as follows
\begin{equation}
\label{eq:simpInt2}
    \itbf{I}^{(3)}_{\operatorname{tr}(\phi^2),3}= \mathcal{I}^{(3),\pl}_{\operatorname{tr}(\phi^2)} + {12 \over N_c^2} \,  \mathcal{I}^{(3),\np}_{\operatorname{tr}(\phi^2)}  \,,
\end{equation}
where the planar (leading $N_c$) and non-planar (sub-leading $N_c$) part can be expanded in terms of integrals respectively as
\begin{align}
        \mathcal{I}^{(3),\pl}_{\operatorname{tr}(\phi^2)} &=  \sum_{i} \sum_{\sigma_{i}}\sigma_i \cdot [ c_i \, {I}_i^{\rm PL} ]\,, 
        \label{eq:simpInt2pl}\\
        \mathcal{I}^{(3),\np}_{\operatorname{tr}(\phi^2)} &=  \sum_{j}\sum_{\sigma_{j}}\sigma_j \cdot [ c_j \, {I}_j^{\rm NP} ]\,.\label{eq:simpInt2np}
\end{align}
Here, $\sigma_{i,j}$ are permutations acting on the three external momenta $\{p_1,p_2,p_3\}$ and are determined by the diagrammatic symmetry, the full set of integrals $\{{I}_i^{\rm PL}, {I}_j^{\rm NP}\}$ are defined explicitly with given topologies and numerators in Appendix~\ref{ap:integrals}, and $c_i$ are corresponding coefficients that only depend on external Mandelstam variables $s_{ij}$, all of which can be found in the ancillary file.

\section{Results}\label{sec:results}

In this section, we consider the integration of the three-loop form factor and study the infrared divergence and finite remainder. 

For numerical calculation, we consider the special kinematic point 
\begin{equation}\label{eq:sij}
s_{12}=s_{23}=s_{13}=-2 \,,
\end{equation}
which possesses the $S_3$ permutational symmetry for the external momenta. This choice of kinematics simplifies the computation by reducing the number of independent integrals, while it also suffices to capture the essential information about IR divergences, as we will discuss below.

We first use the {\tt AMFlow} package \cite{Liu:2022chg},  powered by the block-triangular form improved IBP reduction~\cite{Liu:2018dmc,Guan:2019bcx,www:Blade},  to compute the set of integrals in  \eqref{eq:simpInt2pl} and \eqref{eq:simpInt2np}
 at some given values of $\epsilon$, say $\epsilon= 10^{-4}+ 10^{-6}j$ with $ j=1,2,\cdots,20$. Integrals obtained at this stage have about 60-digit precision. Then, instead of fitting the  $\epsilon$ dependence for each integral, we perform the summations in \eqref{eq:simpInt2pl} and \eqref{eq:simpInt2np} to obtain form factors evaluated at the given values of $\epsilon$, as proposed in Ref.~\cite{Liu:2022mfb}.  Finally, we fit the $\epsilon$-dependence of the form factors based on the above numerical results. The form factors obtained in this way have at least 30-digit precision  up to finite part in $\epsilon$ expansion. In contrast, if one fits the $\epsilon$ dependence of each integral before summing them together, the final results of form factors can only have no more than 25-digit precision due to larger numeric cancellation.

 The full planar and non-planar three-loop form factor results are
\begin{align}
\label{eq:FFPLnum}
\mathcal{I}^{(3),\pl}_{\operatorname{tr}(\phi^2)}  = & 
-\frac{4.50000000000000000000000000000}{\epsilon^6}+\frac{9.35748693755926167713263363969}{\epsilon^5} \nonumber\\
&-\frac{22.6136138848433980268666419286}{\epsilon^4}
   +\frac{55.8893628928813146753414023855}{\epsilon^3}\nonumber\\
   &-\frac{77.2501247502553684012449865425}{\epsilon^2}+\frac{92.9619113279339687879640318517}{\epsilon}\nonumber\\
   &-336.707858549621438955620587295 \,,
\\
\label{eq:FFNPnum}
\mathcal{I}^{(3),\np}_{\operatorname{tr}(\phi^2)}  = & 
-\frac{9.98307291147592432545107273885}{\epsilon}-265.401794126352300525073348339 \,,
\end{align}
where we have truncated all numbers to 30-digit precision.
For reader's convenience, we also provide the $\epsilon$-dependent result of each integral in Appendix~\ref{ap:integrals} and in the ancillary file. To give an idea of the calculation complexity, less than $\mathcal{O}(10^5)$ CPU core hours were used in total, showing its efficiency comparing with the previous Monte-Carlo based numerical calculation in \cite{Lin:2021pne} with nearly $\mathcal{O}(10^7)$ CPU core hours.

Below we analyze their infrared divergences and finite remainders. 

\subsection*{Planar}

We first consider the planar form factor. The three-loop planar $n$-point form factors satisfy the BDS ansatz form \cite{Bern:2005iz} 
\begin{equation}
\label{eq:bds-3loop}
{\mathcal{I}}_n^{(3)}(\epsilon)
= -\frac{1}{3}\left({\mathcal{I}}_n^{(1)}(\epsilon)\right)^{3} + {\mathcal{I}}_n^{(2)}(\epsilon) {\mathcal{I}}_n^{(1)}(\epsilon)+f^{(3)}(\epsilon) {\mathcal{I}}_n^{(1)}(3 \epsilon)+C^{(3)}+\mathcal{R}_n^{(3)}+O(\epsilon)\,,
\end{equation}
where
\begin{equation}\label{eq:f3}
f^{(3)}(\epsilon)=4\Big(\frac{11}{2}\zeta_4 +(6 \zeta_5+5\zeta_2\zeta_3)\epsilon+ X \epsilon^2\Big)\,,
\end{equation}
and both constants $f^{(3)}(\epsilon)$ and $C^{(3)}$ are independent of the number of external legs $n$. We will determine the constant $X$ below.

Using the one and two-loop three-point form factor results \cite{Brandhuber:2010ad, Brandhuber:2012vm}, we find out infrared-divergent parts agree with the BDS ansatz.
At the finite order, the three-loop finite remainder was derived from the bootstrap method \cite{Dixon:2020bbt,Dixon:2021tdw}:
\begin{equation}
{\cal R}_3^{(3)} = -8.3720527811583320031948434406158\,.
\end{equation}
Comparing with our high-precision result at finite order in \eqref{eq:FFPLnum}, we have a relation
\begin{equation}
- \frac{4}{3} X + C^{(3)} = -151.932306781389204458450790649  \,,
\end{equation}
which gives after PSLQ
\begin{equation}
    - \frac{4}{3} X + C^{(3)} = -\frac{76}{3}\zeta_3^2 - \frac{4081}{36} \zeta_6 \,.
\end{equation}

Moreover, one can consider the three-loop Sudakov form factor \cite{Gehrmann:2011xn} and require it to have a zero finite remainder.\footnote{This requirement is analogous to the case of four-point amplitude which has ${\cal R}_{{\rm amp},4}^{(l)} = 0$ \cite{Bern:2005iz}. Equivalently, the condition ${\cal R}_2^{(l)} = 0$ can be interpreted as a way to define $C^{(l)}$. }
From this, we derive another relation:
\begin{equation}
{\cal R}_2^{(3)} = 0 \quad \Rightarrow \quad {11 \pi^6 \over 270} - {8 \over 9} X + C^{(3)} = 8 \left( - {13\over 9} \zeta_3^2 - {193 \over 27} \zeta_6 \right)  \,.
\end{equation}

The above two relations allow us to solve for $X$ and $C^{(3)}$:
\begin{equation}
\label{eq:X}
X = 31 \zeta_3^2 +\frac{1909}{48} \zeta_6 \,, \qquad 
C^{(3)}  =  16 \zeta_3^2 - \frac{181}{3}\zeta_6 \,.
\end{equation}
The value of $X$ is consistent with the numerical result $X = 85.263 \pm 0.004$ of \cite{Spradlin:2008uu} obtained from the computation of three-loop five-point amplitude.
We comment that $X$ is a universal constant for both $n$-point form factors and amplitudes in the BDS ansatz, while the constant $C^{(3)}$ is only for the BDS ansatz of $n$-point form factors \cite{Lin:2021qol}.

\subsection*{Non-planar}

To consider the non-planar form factor, we recall the divergence structure of the full-color three-loop form factor  \cite{Almelid:2015jia} (see \cite{Lin:2021qol} for detailed discussion)
\begin{equation}\label{eq:fullcolorir-3loop}
\begin{aligned}
    \itbf{I}^{(3)}_3=&\frac{1}{6} (\mathfrak{D}^{(1)})^{3} \itbf{I}^{(0)}_3 + \frac{1}{2} (\mathfrak{D}^{(1)})^2 \itbf{I}^{(1),\rm fin}_3 +\mathfrak{D}^{(1)}\itbf{I}^{(2),\rm fin}_3 + \frac{1}{2}\left(\mathfrak{D}^{(2)}\mathfrak{D}^{(1)}+\mathfrak{D}^{(1)}\mathfrak{D}^{(2)}\right)\itbf{I}^{(0)}_3\\
    & + \mathfrak{D}^{(2)}\itbf{I}^{(1),\rm fin}_3 + \mathfrak{D}^{(3)}\itbf{I}^{(0)}_3 +  \frac{1}{3\epsilon} \boldsymbol{\Delta}^{(3)}\itbf{I}^{(0)}_3+\itbf{I}^{(3),\rm fin}_3\,.
\end{aligned}
\end{equation}
Note that the non-planar ($N_c$-subleading) form factor first appears at three loops, and the only source to generate the $N_c$-subleading IR contribution is the non-dipole term $\boldsymbol{\Delta}^{(3)}$, which predicts the following divergence
\begin{equation}\label{eq:non-planarir}
    \mathcal{I}^{(3),\np}_{\operatorname{tr}(\phi^2)} \Big|_{\rm div} = -{1\over\epsilon}(2\zeta_5+4\zeta_2\zeta_3) \,. 
\end{equation}
Our result of the infrared-divergent part in \eqref{eq:FFPLnum} agrees with this result perfectly.

Finally, we present the non-planar three-loop finite part at high precision at the special kinematics \eqref{eq:sij} as 
\begin{equation}
\mathcal{I}^{(3),\np}_{\operatorname{tr}(\phi^2)} \Big|_{\rm fin} = -265.401794126352300525073348339 \,.
\end{equation}

\section{Summary and Outlook}\label{sec:summary}

In this paper, we conduct an efficient high-precision computation of the full-color three-loop three-point form factor of the stress-tensor supermultiplet in ${\cal N}=4$ SYM at a special kinematics point $s_{12}=s_{23}=s_{13}=-2$. We have verified that the infrared-divergent parts agree with previous predictions. Our high-precision calculation of the finite part of the planar form factor helps to determine the analytic form \eqref{eq:X} of the constant $X$ in the BDS ansatz. We also provide high-precision result for the finite part of the non-planar form factor.

The form factor investigated corresponds to a ${\cal N}=4$ version of the Higgs-plus-3-gluon amplitudes in the effective theory where the top quark mass approaches infinity \cite{Ellis:1975ap, Georgi:1977gs, Wilczek:1977zn, Shifman:1979eb, Kniehl:1995tn}. 
In particular, it was previously observed that the two-loop result in ${\cal N}=4$ SYM \cite{Brandhuber:2012vm} equals the maximally transcendental part of the QCD result \cite{Gehrmann:2011aa}, satisfying the so-called principle of maximally transcendentality \cite{Kotikov:2002ab,Kotikov:2004er}. 
Moreover, recent findings reveal a fascinating antipodal duality between the three-loop planar form factor and the six-point amplitude in ${\cal N}=4$ SYM \cite{Dixon:2021tdw}. It remains a compelling question whether this duality holds for the non-planar case as well.
We expect the high-precision results, particularly for the non-planar form factor that firstly appears at three loops, will be valuable for further analytical evaluations.
Additionally, the form factor and three-loop integrals at other kinematics points can be computed efficiently using the differential equation method~\cite{Kotikov:1990kg, Remiddi:1997ny,Henn:2013pwa,Lee:2014ioa,Moriello:2019yhu, Hidding:2020ytt, Armadillo:2022ugh}, where the results of this paper serve as a boundary condition. We leave this for the future study.

\acknowledgments

We would like to thank Siyuan Zhang for collaboration at the early stage of this project. 
This work is supported by the National Natural Science Foundation of China (Grants No.~11935013, 11975029, 12047503, 12175291, 12325503).
We also thank the support of the HPC Cluster of ITP-CAS and the
High-performance Computing Platform of Peking University.

\paragraph{Note added:} While our paper was being finalized, we were acknowledged that  the constant in $f^{(3)}$ has also been calculated using a different method by Chen {et al.} \cite{Chen:2023}. Our results are in full agreement with each other.

\appendix

\section{Definition of integrals}\label{ap:integrals}

In this appendix, we give the definition of integrals appearing in \eqref{eq:simpInt2pl} and \eqref{eq:simpInt2np}.
From the topologies of the integrals one can read propagators. 
We also give the integral results evaluated
at a special kinematics point 
\begin{equation}
s_{12}=s_{23}=s_{13}=-2 \,.
\end{equation} 
We provide the lower-precision results here, while high-precision results (25-digit) can be found in the ancillary file.

A word about convention: we employ the ${\overline {\rm MS}}$ normalization convention for our integrals, for example,
\begin{equation}
I = e^{3 \cdot \gamma_{E}(2-D/2)} \int  \frac{\mathrm{d}\ell_a^{D}}{ \mathrm{i} \pi^{D/2}} \frac{\mathrm{d}\ell_b^{D}}{ \mathrm{i} \pi^{D/2}} \frac{\mathrm{d}\ell_c^{D}}{ \mathrm{i} \pi^{D/2}} \frac{{N}}{D_1^{\nu_1}\dots D_K^{\nu_K}} \,,
\end{equation} 
where $D_{i}$'s are the propagators with the  form $(\ell-p)^2$ and ${N}$ is the irreducible numerator.

\subsection{Planar Integrals}
\begin{equation}
\begin{aligned}
  I^{\rm PL}_{1}  = &  \begin{aligned}
        \includegraphics[width=0.23\linewidth]{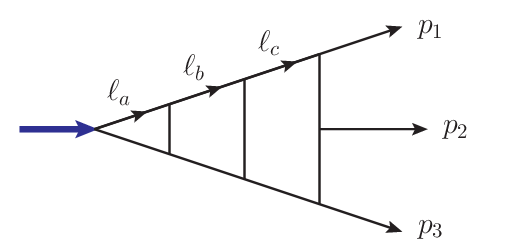}
    \end{aligned}  \\
    = &  +\frac{0.004988702663}{\epsilon^4} - \frac{0.03071310012}{\epsilon^3}  +\frac{0.1081829412}{\epsilon^2} + \frac{0.04865915302}{\epsilon} \\
    &- {1.416843694}\epsilon^0
\end{aligned}
\end{equation}

\begin{equation}
\begin{aligned}
  I^{\rm PL}_{2}  = &  \begin{aligned}
        \includegraphics[width=0.23\linewidth]{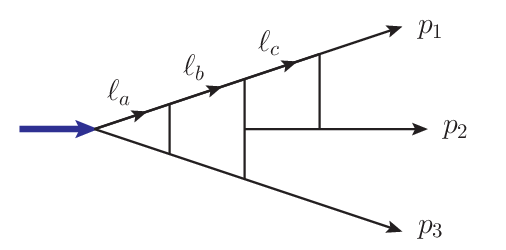}
    \end{aligned}  \times (-\ell_b +p_1)^2\\
    = & - \frac{0.02494351331}{\epsilon^4} + \frac{0.03964337275}{\epsilon^3}  + \frac{0.002503386002}{\epsilon^2} - \frac{0.1992313432}{\epsilon}\\
    & - {3.490313336}\epsilon^0
\end{aligned}
\end{equation}

\begin{equation}
\begin{aligned}
  I^{\rm PL}_{3}  = &  \begin{aligned}
        \includegraphics[width=0.23\linewidth]{figure/phi2Gammaa2.eps}
    \end{aligned}  \\
    = & - \frac{0.01716581701}{\epsilon^5}+ \frac{0.04123213222}{\epsilon^4} - \frac{0.0003871277309}{\epsilon^3}  - \frac{0.2481121613}{\epsilon^2} \\
    & - \frac{1.876408942}{\epsilon}  - {8.200959690}\epsilon^0
\end{aligned}
\end{equation}

\begin{equation}
\begin{aligned}
  I^{\rm PL}_{4}  = &  \begin{aligned}
        \includegraphics[width=0.23\linewidth]{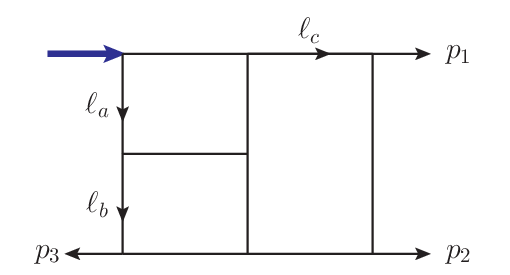}
    \end{aligned} \times \left((-\ell_a+p_1)^2 \right)^2 \\
    = & -\frac{0.08010714605}{\epsilon^5} - \frac{0.2066795960}{\epsilon^4}-\frac{0.6659059362}{\epsilon^3} -\frac{0.8890399741}{\epsilon^2} \\
    &-\frac{1.122171189}{\epsilon}+{5.780193736}\epsilon^0
\end{aligned}
\end{equation}
\begin{equation}
\begin{aligned}
  I^{\rm PL}_{5}  = &  \begin{aligned}
        \includegraphics[width=0.23\linewidth]{figure/phi2Gammaa3.eps}
    \end{aligned} \times(-\ell_a+p_1)^2 \\
    = & -\frac{0.02777777778}{\epsilon^{6}}-\frac{0.03378875901}{\epsilon^5} - \frac{0.1023131258}{\epsilon^4} -\frac{0.4189195604}{\epsilon^3} \\
    &-\frac{2.044946987}{\epsilon^2} -\frac{7.319293288}{\epsilon}-{22.15132708}\epsilon^0
\end{aligned}
\end{equation}
\begin{equation}
\begin{aligned}
  I^{\rm PL}_{6}  = &  \begin{aligned}
        \includegraphics[width=0.23\linewidth]{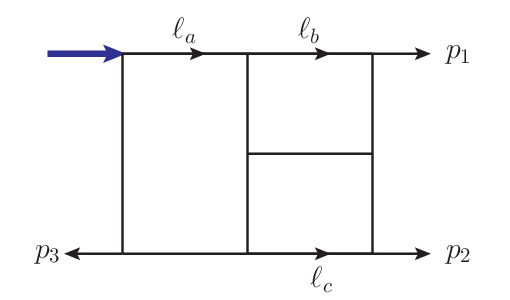}
    \end{aligned} \times(-\ell_a+p_2+p_3)^2(-\ell_c+q)^2 \\
    = & -\frac{0.04577551203}{\epsilon^5} + \frac{0.1299071632}{\epsilon^4} + \frac{0.008952934204}{\epsilon^3}  + \frac{0.2599765423}{\epsilon^2} \\
    &- \frac{1.799167500}{\epsilon}-{3.000778787}\epsilon^0
\end{aligned}
\end{equation}

\begin{equation}
\begin{aligned}
  I^{\rm PL}_{7}  = &  \begin{aligned}
        \includegraphics[width=0.23\linewidth]{figure/phi2Gammaa4.eps}
    \end{aligned} \times(-\ell_a+p_3)^2(-\ell_c+q)^2 \\
    = & -\frac{0.04577551203}{\epsilon^5} + \frac{0.01017829933}{\epsilon^4} - \frac{0.6562139946}{\epsilon^3}  - \frac{1.440971646}{\epsilon^2}\\
    & - \frac{5.345889842}{\epsilon}-{14.37534730}\epsilon^0
\end{aligned}
\end{equation}
\begin{equation}
\begin{aligned}
  I^{\rm PL}_{8}  = &  \begin{aligned}
        \includegraphics[width=0.23\linewidth]{figure/phi2Gammaa4.eps}
    \end{aligned} \times(-\ell_c+q)^2 \\
    = & -\frac{0.02777777778}{\epsilon^6} -\frac{0.03378875901}{\epsilon^5} - \frac{0.2508141179}{\epsilon^4} - \frac{0.5517167205}{\epsilon^3} \\
    & - \frac{0.3550980461}{\epsilon^2} + \frac{4.170236709}{\epsilon} +{27.61952862}\epsilon^0
\end{aligned}
\end{equation}
\begin{equation}
\begin{aligned}
  I^{\rm PL}_{9}  = &  \begin{aligned}
        \includegraphics[width=0.23\linewidth]{figure/phi2Gammaa4.eps}
    \end{aligned} \times(-\ell_a+p_3)^2 \\
    = & +\frac{0.02860969502}{\epsilon^5} - \frac{0.04377670705}{\epsilon^4} + \frac{0.2943548743}{\epsilon^3}  + \frac{1.043173206}{\epsilon^2} \\
    &+ \frac{3.855887202}{\epsilon} +{11.73370532}\epsilon^0
\end{aligned}
\end{equation}

\begin{equation}
\begin{aligned}
  I^{\rm PL}_{10}  = &  \begin{aligned}
        \includegraphics[width=0.23\linewidth]{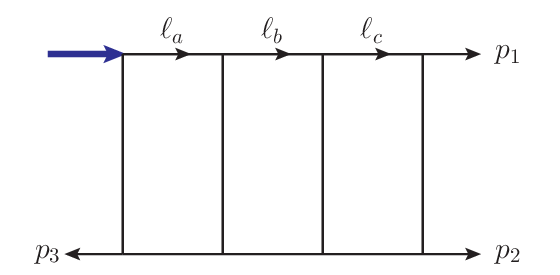}
    \end{aligned}  \times (-\ell_a + p_1)^2\\
    = & + \frac{0.04005357302}{\epsilon^5}  + \frac{0.1033397980}{\epsilon^4} + \frac{0.3827434415}{\epsilon^3}  + \frac{0.05664258922}{\epsilon^2} \\
    &- \frac{2.463136356}{\epsilon} - {17.58032883}\epsilon^0
\end{aligned}
\end{equation}

\begin{equation}
\begin{aligned}
  I^{\rm PL}_{11}  = &  \begin{aligned}
        \includegraphics[width=0.23\linewidth]{figure/phi2Gammaa5.eps}
    \end{aligned}  \\
    = & + \frac{0.004775242360}{\epsilon^3}  - \frac{0.03424056377}{\epsilon^2} + \frac{0.09544237858}{\epsilon} + {0.4966007263}\epsilon^0
\end{aligned}
\end{equation}

\begin{equation}
\begin{aligned}
  I^{\rm PL}_{12}  = &  \begin{aligned}
        \includegraphics[width=0.23\linewidth]{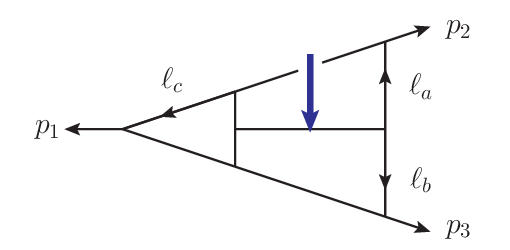}
    \end{aligned} \times(-\ell_a+p_1+p_2)^2 (-(\ell_a+\ell_b+p_2)^2+(\ell_a+\ell_b+p_3)^2) \\
    = & -\frac{0.07291666667}{\epsilon^{6}}-\frac{0.1459148824}{\epsilon^5} + \frac{0.6336506549}{\epsilon^4}+\frac{4.832575843}{\epsilon^3} \\
    &+\frac{15.07737667}{\epsilon^2} +\frac{19.29558033}{\epsilon}{-77.48605638}\epsilon^0
\end{aligned}
\end{equation}

\begin{equation}
\begin{aligned}
  I^{\rm PL}_{13}  = &  \begin{aligned}
        \includegraphics[width=0.23\linewidth]{figure/phi2Gammaa8.eps}
    \end{aligned} \times\left((-\ell_a+p_1+p_2)^2\right)^2 \\
    = & +\frac{0.1006944444}{\epsilon^{6}}-\frac{0.05680317070}{\epsilon^5} - \frac{0.7513807218}{\epsilon^4}-\frac{2.588548811}{\epsilon^3} \\
    &-\frac{3.789611104}{\epsilon^2} -\frac{1.882411740}{\epsilon}+{18.69158831}\epsilon^0
\end{aligned}
\end{equation}

\begin{equation}
\begin{aligned}
  I^{\rm PL}_{14}  = &  \begin{aligned}
        \includegraphics[width=0.23\linewidth]{figure/phi2Gammaa8.eps}
    \end{aligned} \times(-\ell_a+p_1+p_2)^2 (-\ell_b +p_1+p_3)^2\\
    = & +\frac{0.02777777778}{\epsilon^{6}}-\frac{0.1340547851}{\epsilon^5} - \frac{0.2874297578}{\epsilon^4}+\frac{1.033592330}{\epsilon^3} \\
    &+\frac{8.185885076}{\epsilon^2} +\frac{21.88163457}{\epsilon}+{10.19632039}\epsilon^0
\end{aligned}
\end{equation}

\begin{equation}
\begin{aligned}
  I^{\rm PL}_{15}  = &  \begin{aligned}
        \includegraphics[width=0.23\linewidth]{figure/phi2Gammaa8.eps}
    \end{aligned} \times(-\ell_a+p_1+p_2)^2 \\
    = & -\frac{0.07233796296}{\epsilon^{6}}+\frac{0.07031541918}{\epsilon^5} + \frac{0.7331864327}{\epsilon^4}+\frac{1.354941117}{\epsilon^3} \\
    &-\frac{1.918941893}{\epsilon^2} -\frac{15.69832957}{\epsilon}-{38.91023221}\epsilon^0
\end{aligned}
\end{equation}

\begin{equation}
\begin{aligned}
  I^{\rm PL}_{16}  = &  \begin{aligned}
        \includegraphics[width=0.23\linewidth]{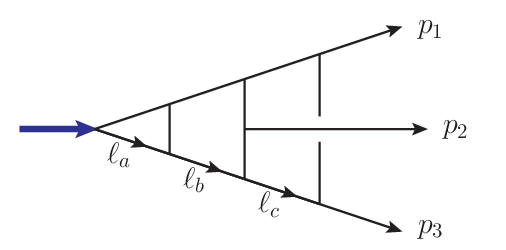}
    \end{aligned} \times (\ell_b -p_3)^2 \\
    = &+  \frac{0.001929012346}{\epsilon^5}  - \frac{0.01466176068}{\epsilon^4} + \frac{0.03160239419}{\epsilon^3}  + \frac{0.2037766572}{\epsilon^2} \\
    &+ \frac{2.710037871}{\epsilon} +{14.44256258}\epsilon^0
\end{aligned}
\end{equation}

\begin{equation}
\begin{aligned}
  I^{\rm PL}_{17}  = &  \begin{aligned}
        \includegraphics[width=0.23\linewidth]{figure/phi2Gammaa11.eps}
    \end{aligned}  \\
    = &  -\frac{0.01337289035}{\epsilon^5}  + \frac{0.07707076746}{\epsilon^4} - \frac{0.07582018903}{\epsilon^3} - \frac{1.356317959}{\epsilon^2} \\
    &- \frac{6.701208997}{\epsilon} - {21.77714211}\epsilon^0
\end{aligned}
\end{equation}

\begin{equation}
\begin{aligned}
  I^{\rm PL}_{18}  = &  \begin{aligned}
        \includegraphics[width=0.23\linewidth]{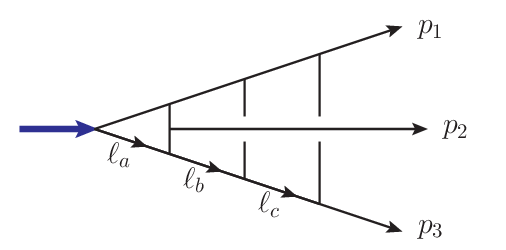}
    \end{aligned} \times(\ell_a - p_3)^2 \\
    = & -\frac{0.004629629630}{\epsilon^6} - \frac{0.007538772836}{\epsilon^5} - \frac{0.05588009571}{\epsilon^4} + \frac{0.3764509608}{\epsilon^3} \\
    & + \frac{2.451868663}{\epsilon^2} + \frac{13.84622030}{\epsilon} +{60.55972130}\epsilon^0
\end{aligned}
\end{equation}

\begin{equation}
\begin{aligned}
  I^{\rm PL}_{19}  = &  \begin{aligned}
        \includegraphics[width=0.23\linewidth]{figure/phi2Gammaa12.eps}
    \end{aligned} \times(\ell_b - p_3)^2 \\
    = & - \frac{0.05028954003}{\epsilon^4} - \frac{0.08694690459}{\epsilon^3}  + \frac{0.07321408103}{\epsilon^2} + \frac{3.502206610}{\epsilon} \\
    &+{23.03684664}\epsilon^0
\end{aligned}
\end{equation}

\begin{equation}
\begin{aligned}
  I^{\rm PL}_{20}  = &  \begin{aligned}
        \includegraphics[width=0.23\linewidth]{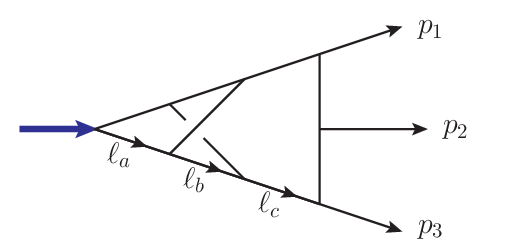}
    \end{aligned}  \\
    = & + \frac{0.01257238501}{\epsilon^4} - \frac{0.03929063174}{\epsilon^3}  - \frac{0.05375749458}{\epsilon^2} - \frac{0.8235364847}{\epsilon} \\
    &- {7.151137688}\epsilon^0
\end{aligned}
\end{equation}

\begin{equation}
\begin{aligned}
  I^{\rm PL}_{21}  = &  \begin{aligned}
        \includegraphics[width=0.23\linewidth]{figure/phi2Gammaa12.eps}
    \end{aligned}  \\
    = & + \frac{0.01504629630}{\epsilon^6}  - \frac{0.01984401556}{\epsilon^5}  - \frac{0.1638550577}{\epsilon^4} - \frac{0.7550462381}{\epsilon^3} \\
    & - \frac{2.020128579}{\epsilon^2} - \frac{3.252063922}{\epsilon} +{1.411788137}\epsilon^0
\end{aligned}
\end{equation}

\begin{equation}
\begin{aligned}
  I^{\rm PL}_{22}  = &  \begin{aligned}
        \includegraphics[width=0.23\linewidth]{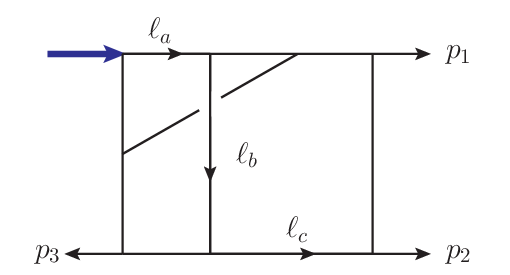}
    \end{aligned} \times(2\ell_a \cdot p_2)(\ell_c+p_3 )^2\\
    = &+ \frac{0.04976851852}{\epsilon^{6}}- \frac{0.08754741331}{\epsilon^5} - \frac{0.1414492084}{\epsilon^4}+\frac{0.3143018346}{\epsilon^3} \\
    &+\frac{5.663722382}{\epsilon^2} +\frac{33.07718694}{\epsilon}+{148.3820735}\epsilon^0
\end{aligned}
\end{equation}

\begin{equation}
\begin{aligned}
  I^{\rm PL}_{23}  = &  \begin{aligned}
        \includegraphics[width=0.23\linewidth]{figure/phi2Gammaa15.eps}
    \end{aligned} \times(2\ell_a \cdot p_1)(\ell_c+p_3 )^2\\
    = & +\frac{0.07060185185}{\epsilon^{6}}- \frac{0.05076196604}{\epsilon^5} - \frac{0.1154706956}{\epsilon^4}+\frac{0.1311860493}{\epsilon^3} \\
    &+\frac{2.848775690}{\epsilon^2} +\frac{10.30897990}{\epsilon}+{19.89793516}\epsilon^0
\end{aligned}
\end{equation}

\begin{equation}
\begin{aligned}
  I^{\rm PL}_{24}  = &  \begin{aligned}
        \includegraphics[width=0.23\linewidth]{figure/phi2Gammaa15.eps}
    \end{aligned} \times(\ell_b-\ell_c+p_2)^2\\
    = & -\frac{0.003858024691}{\epsilon^{6}}+ \frac{0.02466671742}{\epsilon^5} + \frac{0.02214993866}{\epsilon^4}+\frac{0.008602078935}{\epsilon^3} \\
    &-\frac{1.425807633}{\epsilon^2} -\frac{11.83611109}{\epsilon}-{66.19054263}\epsilon^0
\end{aligned}
\end{equation}

\begin{equation}
\begin{aligned}
  I^{\rm PL}_{25}  = &  \begin{aligned}
        \includegraphics[width=0.23\linewidth]{figure/phi2Gammaa15.eps}
    \end{aligned} \times(\ell_a + \ell_c -p_1 -p_2)^2\\
    = &+ \frac{0.005787037037}{\epsilon^{6}}+ \frac{0.05669456086}{\epsilon^5} - \frac{0.01087355968}{\epsilon^4}+\frac{0.5155274959}{\epsilon^3} \\
    &+\frac{3.153098447}{\epsilon^2} +\frac{19.31223582}{\epsilon}+{93.23578735}\epsilon^0
\end{aligned}
\end{equation}

\begin{equation}
\begin{aligned}
  I^{\rm PL}_{26}  = &  \begin{aligned}
        \includegraphics[width=0.23\linewidth]{figure/phi2Gammaa15.eps}
    \end{aligned} \times(\ell_a -p_1 -p_2)^2\\
    = & -\frac{0.001157407407}{\epsilon^{6}}+ \frac{0.01010111108}{\epsilon^5} - \frac{0.07986297156}{\epsilon^4}+\frac{0.5968937704}{\epsilon^3} \\
    &+\frac{4.554087484}{\epsilon^2} +\frac{25.34381039}{\epsilon}+{111.1403172}\epsilon^0
\end{aligned}
\end{equation}

\begin{equation}
\begin{aligned}
  I^{\rm PL}_{27}  = &  \begin{aligned}
        \includegraphics[width=0.23\linewidth]{figure/phi2Gammaa15.eps}
    \end{aligned} \\
    = &+ \frac{0.01552854938}{\epsilon^{6}}- \frac{0.001031827793}{\epsilon^5} + \frac{0.02461164706}{\epsilon^4}-\frac{0.003959251408}{\epsilon^3} \\
    &-\frac{0.2058633961}{\epsilon^2} +\frac{0.2359249558}{\epsilon}+{8.671670255}\epsilon^0
\end{aligned}
\end{equation}

\begin{equation}
\begin{aligned}
  I^{\rm PL}_{28}  = &  \begin{aligned}
        \includegraphics[width=0.23\linewidth]{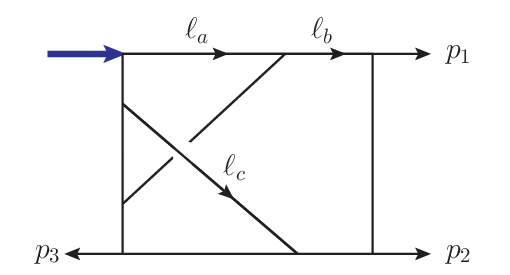}
    \end{aligned} \times (\ell_a - p_1 - p_2)^2 (-\ell_b + q)^2\\
    = & - \frac{0.01851851852}{\epsilon^{6}}- \frac{0.1255207414}{\epsilon^5} - \frac{0.1168251380}{\epsilon^4}-\frac{0.7242311771}{\epsilon^3} \\
    &-\frac{0.6223046326}{\epsilon^2} +\frac{3.082184410}{\epsilon}+{42.36931238}\epsilon^0
\end{aligned}
\end{equation}

\begin{equation}
\begin{aligned}
  I^{\rm PL}_{29}  = &  \begin{aligned}
        \includegraphics[width=0.23\linewidth]{figure/phi2Gammaa19.eps}
    \end{aligned} \times (\ell_a - p_1)^2 (-\ell_b + q)^2\\
    = &+ \frac{0.002314814815}{\epsilon^{6}}- \frac{0.03151590410}{\epsilon^5} + \frac{0.1145616837}{\epsilon^4}-\frac{0.2636683315}{\epsilon^3} \\
    &-\frac{0.02023532580}{\epsilon^2} +\frac{2.533234394}{\epsilon}+{29.69122141}\epsilon^0
\end{aligned}
\end{equation}

\begin{equation}
\begin{aligned}
  I^{\rm PL}_{30}  = &  \begin{aligned}
        \includegraphics[width=0.23\linewidth]{figure/phi2Gammaa19.eps}
    \end{aligned} \times(-\ell_b + q)^2\\
    = & -\frac{0.03877314815}{\epsilon^{6}}- \frac{0.007109903092}{\epsilon^5} - \frac{0.01887211423}{\epsilon^4}+\frac{0.1701131712}{\epsilon^3} \\
    &+\frac{1.415865336}{\epsilon^2} +\frac{9.181182283}{\epsilon}+{45.28920780}\epsilon^0
\end{aligned}
\end{equation}

\begin{equation}
\begin{aligned}
  I^{\rm PL}_{31}  = &  \begin{aligned}
        \includegraphics[width=0.23\linewidth]{figure/phi2Gammaa19.eps}
    \end{aligned} \times(\ell_a - p_1 - p_2)^2\\
    = & +\frac{0.01118827160}{\epsilon^{6}}+ \frac{0.05985747314}{\epsilon^5} + \frac{0.007472764094}{\epsilon^4}+\frac{0.5263024061}{\epsilon^3} \\
    &+\frac{2.415074702}{\epsilon^2} +\frac{8.746508645}{\epsilon}+{19.72226764}\epsilon^0
\end{aligned}
\end{equation}

\begin{equation}
\begin{aligned}
  I^{\rm PL}_{32}  = &  \begin{aligned}
        \includegraphics[width=0.23\linewidth]{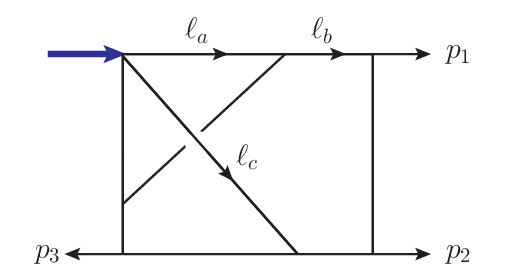}
    \end{aligned} \times (\ell_c+p_1)^2 \\
    = &+ \frac{0.05902777778}{\epsilon^{6}}- \frac{0.1456325692}{\epsilon^5} - \frac{0.2027792134}{\epsilon^4}-\frac{0.6846801064}{\epsilon^3} \\
    &+\frac{1.565265806}{\epsilon^2} +\frac{9.399529667}{\epsilon}+{38.18702969}\epsilon^0
\end{aligned}
\end{equation}

\begin{equation}
\begin{aligned}
  I^{\rm PL}_{33}  = &  \begin{aligned}
        \includegraphics[width=0.23\linewidth]{figure/phi2Gammaa8.eps}
    \end{aligned} \times (\ell_a - p_2)^2 \\
    = & -\frac{0.0003858024691}{\epsilon^6}  + \frac{0.005660152498}{\epsilon^5}  - \frac{0.01471868858}{\epsilon^4} - \frac{0.02010201047}{\epsilon^3} \\
    = & -\frac{ 0.01223558547}{\epsilon^2} + \frac{ 1.068024779}{\epsilon} + 8.0805715247 \epsilon^0
\end{aligned}
\end{equation}

\begin{equation}
\begin{aligned}
  I^{\rm PL}_{34}  = &  \begin{aligned}
        \includegraphics[width=0.23\linewidth]{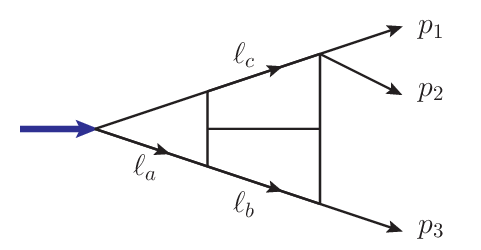}
    \end{aligned}  \times (\ell_a- p_3)^2\\
    = &   - \frac{0.01158530114}{\epsilon^3}  +\frac{0.07219656483}{\epsilon^2}  - \frac{0.2498542316}{\epsilon} - {0.1239782660}\epsilon^0
\end{aligned}
\end{equation}

\begin{equation}
\begin{aligned}
  I^{\rm PL}_{35}  = &  \begin{aligned}
        \includegraphics[width=0.23\linewidth]{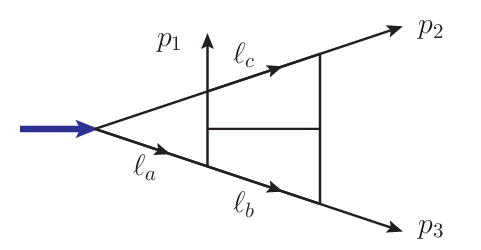}
    \end{aligned}  \times (\ell_a - p_3)^2\\
    = &   -\frac{0.01144387801}{\epsilon^5}  +\frac{0.07737511477}{\epsilon^4}  - \frac{0.06541086481}{\epsilon^3}  -\frac{0.1908458531}{\epsilon^2}  \\
    &- \frac{2.487935703}{\epsilon} - {11.78273194}\epsilon^0
\end{aligned}
\end{equation}
\begin{equation}
\begin{aligned}
  I^{\rm PL}_{36}  = &  \begin{aligned}
        \includegraphics[width=0.23\linewidth]{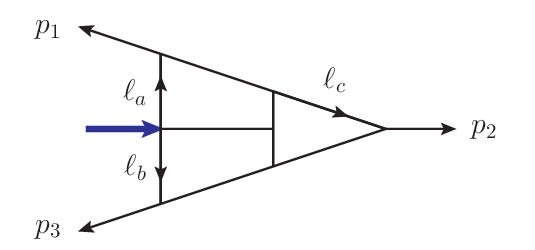}
    \end{aligned}  \times (\ell_b - p_2 - p_3)^2\\
    = & +  \frac{0.04861111111}{\epsilon^6}   -\frac{0.1010839638}{\epsilon^5}  - \frac{0.3425098100}{\epsilon^4}  - \frac{0.9284540918}{\epsilon^3} \\
    & -\frac{0.7016321430}{\epsilon^2}  - \frac{2.161180083}{\epsilon} - {14.51695239}\epsilon^0
\end{aligned}
\end{equation}

\begin{equation}
\begin{aligned}
  I^{\rm PL}_{37}  = &  \begin{aligned}
        \includegraphics[width=0.23\linewidth]{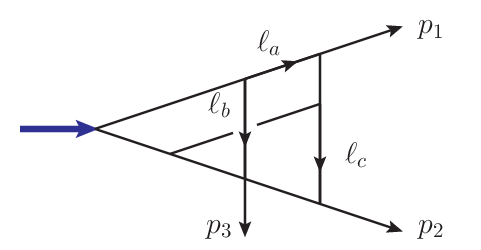}
    \end{aligned}  \times (-\ell_b+p_2+p_3)^2\\
    = &    -\frac{0.02288775601}{\epsilon^5}  + \frac{0.1287104831}{\epsilon^4}  + \frac{0.05003666587}{\epsilon^3}  -\frac{0.2998697024}{\epsilon^2}  \\
    &- \frac{3.187512584}{\epsilon} - {7.913302690}\epsilon^0
\end{aligned}
\end{equation}

\begin{equation}
\begin{aligned}
  I^{\rm PL}_{38}  = &  \begin{aligned}
        \includegraphics[width=0.23\linewidth]{figure/phi2Gamma94.eps}
    \end{aligned}  \times (-\ell_a+p_1+p_2)^2\\
    = &  +  \frac{0.03771715503}{\epsilon^4}  - \frac{0.1223564048}{\epsilon^3}  -\frac{0.3750614030}{\epsilon^2} +\frac{1.211044132}{\epsilon}\\
    & + {19.60355026}\epsilon^0
\end{aligned}
\end{equation}

\begin{equation}
\begin{aligned}
  I^{\rm PL}_{39}  = &  \begin{aligned}
        \includegraphics[width=0.23\linewidth]{figure/phi2Gamma94.eps}
    \end{aligned}  \\
    = &+ \frac{0.01273148148}{\epsilon^5} - \frac{0.06226964468}{\epsilon^4} - \frac{0.08095467186}{\epsilon^3}  +\frac{0.2313400069}{\epsilon^2} \\
    & + \frac{0.7762286087}{\epsilon} - {3.806749978}\epsilon^0
\end{aligned}
\end{equation}

\begin{equation}
\begin{aligned}
  I^{\rm PL}_{40}  = &  \begin{aligned}
        \includegraphics[width=0.23\linewidth]{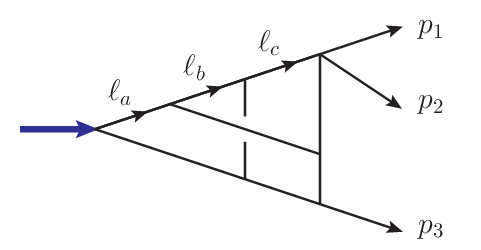}
    \end{aligned}  \times (-\ell_a+p_1+p_2)^2\\
    = &   - \frac{0.005146653277}{\epsilon^3}  +\frac{0.06608582577}{\epsilon^2} - \frac{0.2582690853}{\epsilon} + {0.05888629270}\epsilon^0
\end{aligned}
\end{equation}

\begin{equation}
\begin{aligned}
  I^{\rm PL}_{41}  = &  \begin{aligned}
        \includegraphics[width=0.23\linewidth]{figure/phi2Gamma97.eps}
    \end{aligned}  \times (-\ell_b+q)^2\\
    = & +\frac{0.05902777778}{\epsilon^6}  -\frac{0.1456325692}{\epsilon^5}  - \frac{0.2027792134}{\epsilon^4}  - \frac{0.6846801064}{\epsilon^3} \\
    & +\frac{1.565265806}{\epsilon^2} + \frac{9.399529667}{\epsilon} + {38.18702969}\epsilon^0
\end{aligned}
\end{equation}

\begin{equation}
\begin{aligned}
  I^{\rm PL}_{42}  = &  \begin{aligned}
        \includegraphics[width=0.23\linewidth]{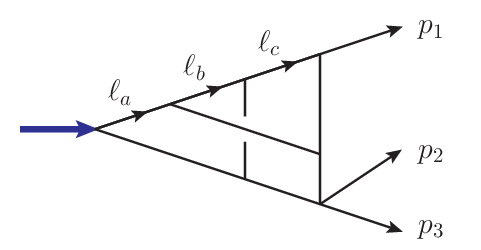}
    \end{aligned}  \times (-\ell_a+p_1)^2\\
    = &  - \frac{0.01827694121}{\epsilon^3}  +\frac{0.09564147142}{\epsilon^2} - \frac{0.1670869169}{\epsilon} - {0.2744493177}\epsilon^0
\end{aligned}
\end{equation}

\begin{equation}
\begin{aligned}
  I^{\rm PL}_{43}  = &  \begin{aligned}
        \includegraphics[width=0.23\linewidth]{figure/phi2Gammaa11.eps}
    \end{aligned} \times (\ell_c - p_3)^2 \\
    = & +\frac{0.004775242360}{\epsilon^3}  - \frac{0.03424056377}{\epsilon^2}  + \frac{0.09544237858}{\epsilon} + {0.4966007263}\epsilon^0
\end{aligned}
\end{equation}

\begin{equation}
\begin{aligned}
  I^{\rm PL}_{44}  = &  \begin{aligned}
        \includegraphics[width=0.23\linewidth]{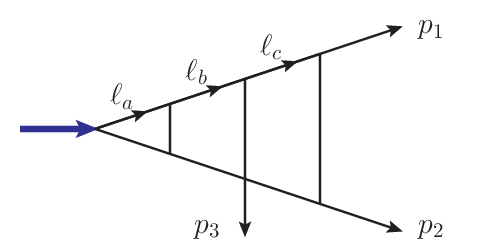}
    \end{aligned}  \\
    = & - \frac{0.004988702663}{\epsilon^4} + \frac{0.03457486717}{\epsilon^3}  -\frac{0.1348547869}{\epsilon^2}  + \frac{0.04983242299}{\epsilon}\\
    & + {1.399389147}\epsilon^0
\end{aligned}
\end{equation}

\begin{equation}
\begin{aligned}
  I^{\rm PL}_{45}  = &  \begin{aligned}
        \includegraphics[width=0.23\linewidth]{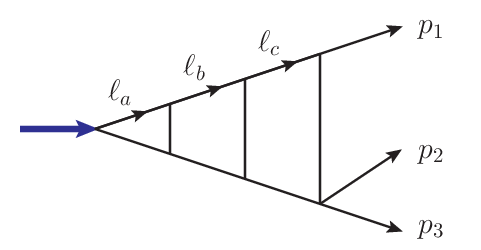}
    \end{aligned}  \\
    = & +\frac{0.001930883523}{\epsilon^3}  -\frac{0.01333592284}{\epsilon^2}  + \frac{0.04924578801}{\epsilon} - {0.03401712147}\epsilon^0
\end{aligned}
\end{equation}

\begin{equation}
\begin{aligned}
  I^{\rm PL}_{46}  = &  \begin{aligned}
        \includegraphics[width=0.23\linewidth]{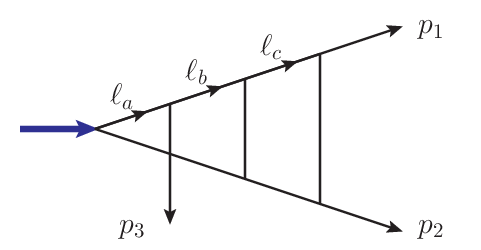}
    \end{aligned}  \\
    = & +\frac{0.005721939003}{\epsilon^5}  - \frac{0.03868755739}{\epsilon^4}  +\frac{0.08917870678}{\epsilon^3}  -\frac{0.05273234781}{\epsilon^2}\\
    &  + \frac{0.2925879960}{\epsilon} + {1.786770094}\epsilon^0
\end{aligned}
\end{equation}

\begin{equation}
\begin{aligned}
  I^{\rm PL}_{47}  = &  \begin{aligned}
        \includegraphics[width=0.23\linewidth]{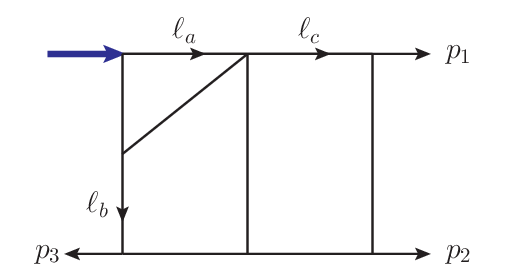}
    \end{aligned}  \\
    = & +\frac{0.02860969502}{\epsilon^5}  - \frac{0.1036411390}{\epsilon^4}  +\frac{0.06003202853}{\epsilon^3} +\frac{0.4890951224}{\epsilon^2} \\
    & + \frac{2.819199089}{\epsilon} + {8.112944470}\epsilon^0
\end{aligned}
\end{equation}

\begin{equation}
\begin{aligned}
  I^{\rm PL}_{48}  = &  \begin{aligned}
        \includegraphics[width=0.23\linewidth]{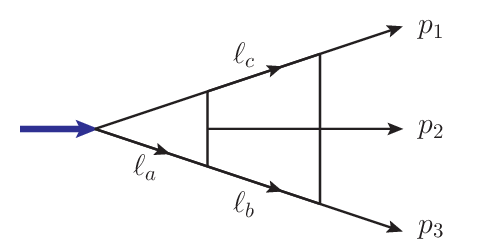}
    \end{aligned}  \\
    = &+ \frac{0.1335618781}{\epsilon^3}  +\frac{0.3983705037}{\epsilon^2}  + \frac{1.179393553}{\epsilon} + {4.179065123}\epsilon^0
\end{aligned}
\end{equation}

\begin{equation}
\begin{aligned}
  I^{\rm PL}_{49}  = &  \begin{aligned}
        \includegraphics[width=0.23\linewidth]{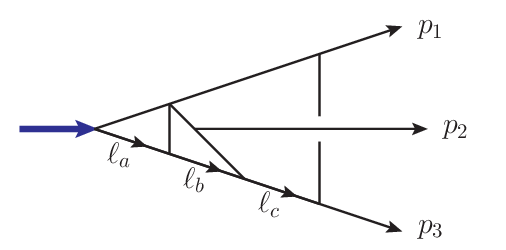}
    \end{aligned}\\
    = &   + \frac{0.01723091504}{\epsilon^5} - \frac{0.1163716941}{\epsilon^4} + \frac{0.1918141683}{\epsilon^3}  +\frac{1.416400313}{\epsilon^2} \\
    & + \frac{6.074123542}{\epsilon} + {15.66192969}\epsilon^0
\end{aligned}
\end{equation}

\begin{equation}
\begin{aligned}
  I^{\rm PL}_{50}  = &  \begin{aligned}
        \includegraphics[width=0.23\linewidth]{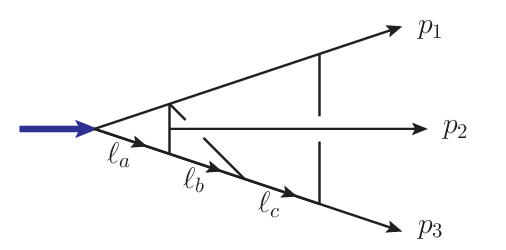}
    \end{aligned}\\
    = &   -  \frac{0.004629629630}{\epsilon^6} +  \frac{0.03251480019}{\epsilon^5} +  \frac{0.02470865525}{\epsilon^4} - \frac{0.08480578859}{\epsilon^3} \\
    & - \frac{1.692826141}{\epsilon^2}  - \frac{9.016453488}{\epsilon} - {36.95902603}\epsilon^0
\end{aligned}
\end{equation}

\begin{equation}
\begin{aligned}
  I^{\rm PL}_{51}  = &  \begin{aligned}
        \includegraphics[width=0.23\linewidth]{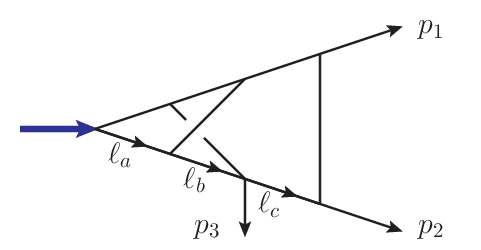}
    \end{aligned}\\
    = &   -  \frac{0.01257238501}{\epsilon^4} + \frac{0.04816789391}{\epsilon^3} + \frac{0.01966586705}{\epsilon^2}  + \frac{0.8039962966}{\epsilon}\\
    & + {6.578202759}\epsilon^0
\end{aligned}
\end{equation}

\begin{equation}
\begin{aligned}
  I^{\rm PL}_{52}  = &  \begin{aligned}
        \includegraphics[width=0.23\linewidth]{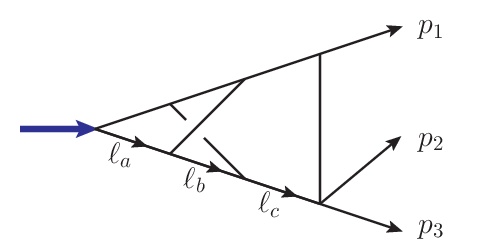}
    \end{aligned}\\
    = & +   \frac{0.004438631087}{\epsilon^3}  - \frac{0.01704581376}{\epsilon^2}  - \frac{0.009770094009}{\epsilon} - {0.2546359830}\epsilon^0
\end{aligned}
\end{equation}

\begin{equation}
\begin{aligned}
  I^{\rm PL}_{53}  = &  \begin{aligned}
        \includegraphics[width=0.23\linewidth]{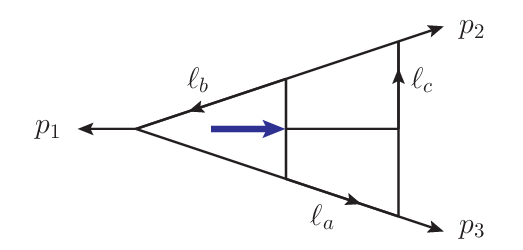}
    \end{aligned} \times (\ell_a-p_2-p_3)^2 \\
    = & \frac{0.1562500000}{\epsilon^6} + \frac{0.2333617168}{\epsilon^5} - \frac{1.420241193}{\epsilon^4} - \frac{1.957240838}{\epsilon^3} \\
    & + \frac{8.026361751}{\epsilon^2}  +  \frac{44.51461807}{\epsilon} + {119.2673216}\epsilon^0
\end{aligned}
\end{equation}

\begin{equation}
\begin{aligned}
  I^{\rm PL}_{54}  = &  \begin{aligned}
        \includegraphics[width=0.23\linewidth]{figure/phi2Gamma95.eps}
    \end{aligned}  \\
    = & - \frac{0.1562500000}{\epsilon^6} + \frac{0.1875862048}{\epsilon^5} + \frac{1.701456323}{\epsilon^4} + \frac{2.174091418}{\epsilon^3} \\
    & -\frac{10.81378498}{\epsilon^2}  - \frac{61.40527517}{\epsilon} - {154.8608211}\epsilon^0
\end{aligned}
\end{equation}

\begin{equation}
\begin{aligned}
  I^{\rm PL}_{55}  = &  \begin{aligned}
        \includegraphics[width=0.23\linewidth]{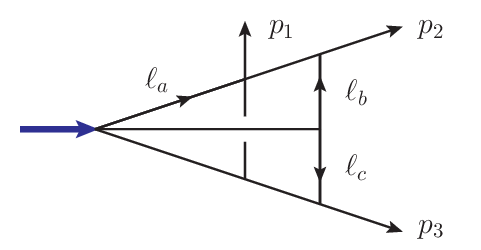}
    \end{aligned}  \\
    = &  - \frac{0.02288775601}{\epsilon^5}  + \frac{0.1790000231}{\epsilon^4} - \frac{0.1131052072}{\epsilon^3}  -\frac{1.474692214}{\epsilon^2}\\
    &- \frac{5.751773424}{\epsilon} - {3.514390529}\epsilon^0
\end{aligned}
\end{equation}

\begin{equation}
\begin{aligned}
  I^{\rm PL}_{56}  = &  \begin{aligned}
        \includegraphics[width=0.23\linewidth]{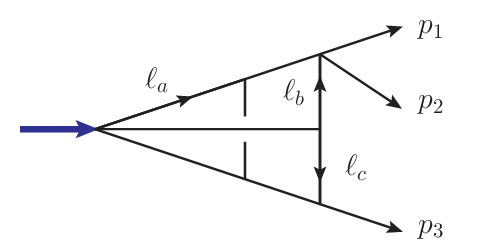}
    \end{aligned}  \\
    = &  - \frac{0.04932585045}{\epsilon^3}  +\frac{0.1581658299}{\epsilon^2}  + \frac{0.07452886439}{\epsilon} + {0.3927644977}\epsilon^0
\end{aligned}
\end{equation}

\subsection{Non-planar Integrals}

\begin{equation}
\begin{aligned}
    I^{\rm NP}_{1}=&\begin{aligned}
    \includegraphics[width=0.23\linewidth]{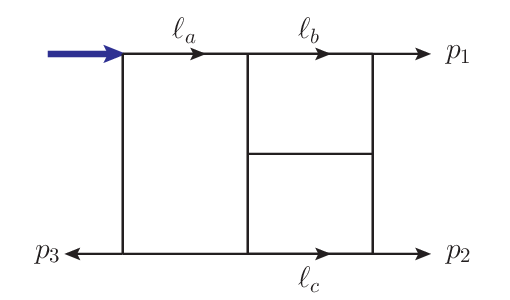}
    \end{aligned} \times  \left((\ell_1-p_1)^2 \right)^2 \\
    =&-\frac{0.08010714605}{\epsilon \
^5}-\frac{0.2066795960}{\epsilon ^4}-\frac{0.6659059362}{\epsilon \
^3} -\frac{0.8890399741}{\epsilon ^2} \\
    &-\frac{1.122171189}{\epsilon
}+5.780193736
\end{aligned}
\end{equation}

\begin{equation}
\begin{aligned}
    I^{\rm NP}_{2}=&\begin{aligned}
    \includegraphics[width=0.23\linewidth]{figure/phi2NplGamma10to1.eps}
    \end{aligned} \times  (\ell_1-p_1)^2  \\
    =&-\frac{0.02777777778}{\epsilon \
^6}-\frac{0.03378875901}{\epsilon ^5}-\frac{0.1023131258}{\epsilon \
^4}-\frac{0.4189195604}{\epsilon
^3}\\
&-\frac{2.044946987}{\epsilon ^2}-\frac{7.319293288}{\epsilon }-22.15132708
\end{aligned}
\end{equation}

\begin{equation}
\begin{aligned}
    I^{\rm NP}_{3}=&\begin{aligned}
    \includegraphics[width=0.23\linewidth]{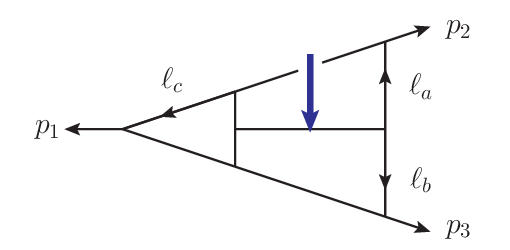}
    \end{aligned} \times  \left((\ell_b-p_1-p_3)^2 \right)^2  \\
    =&+\frac{0.1006944444}{\epsilon \
^6}-\frac{0.05680317070}{\epsilon ^5}-\frac{0.7513807218}{\epsilon \
^4}-\frac{2.588548811}{\epsilon ^3}\\
&-\frac{3.789611104}{\epsilon
^2}-\frac{1.882411740}{\epsilon }+18.69158831
\end{aligned}
\end{equation}

\begin{equation}
\begin{aligned}
    I^{\rm NP}_{4}=&\begin{aligned}
    \includegraphics[width=0.23\linewidth]{figure/phi2NplGamma10to2.eps}
    \end{aligned} \times  (\ell_b-p_1-p_3)^2 (\ell_a-p_1-p_2)^2  \\
   = &+\frac{0.02777777778}{\epsilon \
^6}-\frac{0.1340547851}{\epsilon ^5}-\frac{0.2874297578}{\epsilon \
^4}+\frac{1.033592330}{\epsilon ^3}\\
&+\frac{8.185885076}{\epsilon
^2}+\frac{21.88163457}{\epsilon }+10.19632039
\end{aligned}
\end{equation}

\begin{equation}
\begin{aligned}
    I^{\rm NP}_{5}=&\begin{aligned}
    \includegraphics[width=0.23\linewidth]{figure/phi2NplGamma10to2.eps}
    \end{aligned} \times  (\ell_a-p_1-p_2)^2  \left((\ell_a+\ell_b-p_2)^2 - (\ell_a+\ell_b-p_3)^2\right) \\
    =&+\frac{0.07291666667}{\epsilon \
^6}+\frac{0.1459148824}{\epsilon ^5}-\frac{0.6336506549}{\epsilon \
^4}-\frac{4.832575843}{\epsilon ^3}\\
&-\frac{15.07737667}{\epsilon
^2}-\frac{19.29558033}{\epsilon }+77.48605638
\end{aligned}
\end{equation}

\begin{equation}
\begin{aligned}
    I^{\rm NP}_{6}=&\begin{aligned}
    \includegraphics[width=0.23\linewidth]{figure/phi2NplGamma10to2.eps}
    \end{aligned} \times  (\ell_b-p_1-p_3)^2  \\
    =&-\frac{0.07233796296}{\epsilon \
^6}+\frac{0.07031541918}{\epsilon ^5}+\frac{0.7331864327}{\epsilon \
^4}+\frac{1.354941117}{\epsilon ^3}\\
&-\frac{1.918941893}{\epsilon
^2}-\frac{15.69832957}{\epsilon }-38.91023221
\end{aligned}
\end{equation}

\begin{equation}
\begin{aligned}
    I^{\rm NP}_{7}=&\begin{aligned}
        \includegraphics[width=0.23\linewidth]{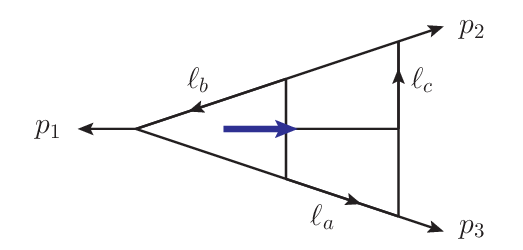}
    \end{aligned} \\
    =&-\frac{0.1562500000}{\epsilon \
^6}+\frac{0.1875862048}{\epsilon ^5}+\frac{1.701456323}{\epsilon ^4}+\
\frac{2.174091418}{\epsilon ^3}\\
&-\frac{10.81378498}{\epsilon
^2}-\frac{61.40527517}{\epsilon }-154.8608211
\end{aligned}
\end{equation}

\begin{equation}
\begin{aligned}
    I^{\rm NP}_{8}=&\begin{aligned}
        \includegraphics[width=0.23\linewidth]{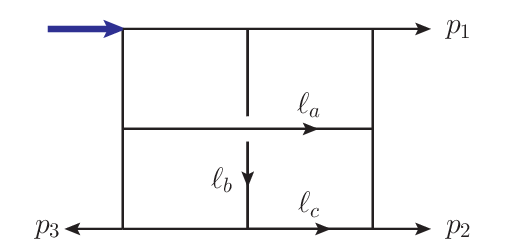}
    \end{aligned} \times \ell_c\cdot (p_3-p_1)\\
    =&  - \frac{0.001041666667}{\epsilon^6}  - \frac{0.003174391464}{\epsilon^5} - \frac{0.05543835820}{\epsilon^4} - \frac{0.3096631584}{\epsilon^3} \\
    & -\frac{2.208464307}{\epsilon^2}  - \frac{17.77926290}{\epsilon} - {129.1610634}\epsilon^0
\end{aligned}
\end{equation}

\begin{equation}
\begin{aligned}
    I^{\rm NP}_{9}=&\begin{aligned}
        \includegraphics[width=0.23\linewidth]{figure/phi2NplGamma10to3.eps}
    \end{aligned} \times \ell_c^2\\
    =& -\frac{0.002893518519}{\epsilon ^6}-\frac{0.005426975398}{\
\epsilon ^5}-\frac{0.05387882213}{\epsilon ^4}+\frac{0.1317873704}{\epsilon
^3} \\
& +\frac{1.751950520}{\epsilon ^2}+\frac{11.26599683}{\epsilon } +62.79642351
\end{aligned}
\end{equation}

\begin{equation}
\begin{aligned}
    I^{\rm NP}_{10}=&\begin{aligned}
        \includegraphics[width=0.23\linewidth]{figure/phi2NplGamma10to3.eps}
    \end{aligned} \times \ell_a\cdot \ell_b\\
    =& -\frac{0.01967592593}{\epsilon ^6}+\frac{0.04701833935}{\epsilon ^5}+\
\frac{0.4140115419}{\epsilon ^4}+\frac{0.5196196662}{\epsilon
^3} \\
& -\frac{4.661749871}{\epsilon ^2}-\frac{41.66687534}{\epsilon } -266.6468569
\end{aligned}
\end{equation}

\begin{equation}
\begin{aligned}
    I^{\rm NP}_{11}=&\begin{aligned}
        \includegraphics[width=0.23\linewidth]{figure/phi2NplGamma10to3.eps}
    \end{aligned} \times \ell_b^2 (\ell_a - \ell_b + \ell_c - p_2)^2\\
   = &-\frac{0.005864197531}{\epsilon ^6}+\frac{0.08071215818}{\epsilon \
^5}-\frac{0.05622457258}{\epsilon ^4}-\frac{1.534405980}{\epsilon
^3} \\
& -\frac{1.120068385}{\epsilon ^2}-\frac{90.92661214}{\epsilon }+333.5313673
\end{aligned}
\end{equation}

\begin{equation}
\begin{aligned}
    I^{\rm NP}_{12}=&\begin{aligned}
        \includegraphics[width=0.23\linewidth]{figure/phi2NplGamma10to3.eps}
    \end{aligned} \times \ell_a^2  (\ell_a - \ell_b + \ell_c - p_2)^2\\
   = &-\frac{0.007407407407}{\epsilon ^6}+\frac{0.02519110222}{\
\epsilon ^5}+\frac{0.2508718932}{\epsilon \
^4}-\frac{0.1596125948}{\epsilon
^3} \\ 
& -\frac{2.425299297}{\epsilon ^2}-\frac{12.64106249}{\epsilon }-53.87113054
\end{aligned}
\end{equation}

\begin{equation}
\begin{aligned}
    I^{\rm NP}_{13}=&\begin{aligned}
        \includegraphics[width=0.23\linewidth]{figure/phi2NplGamma10to3.eps}
    \end{aligned} \times (\ell_a - \ell_b + \ell_c - p_2)^2 \, (-\ell_c + p_2) \cdot (\ell_a - \ell_b + \ell_c + p_3)\\
    =&-\frac{0.007407407407}{\epsilon ^6}+\frac{0.004474945345}{\
\epsilon ^5}+\frac{0.2319985090}{\epsilon \
^4}+\frac{0.5193774107}{\epsilon
^3}\\ 
& +\frac{0.6214427520}{\epsilon ^2}+\frac{2.337752431}{\epsilon }+22.17173713
\end{aligned}
\end{equation}

\begin{equation}
\begin{aligned}
    I^{\rm NP}_{14}=&\begin{aligned}
        \includegraphics[width=0.23\linewidth]{figure/phi2NplGamma10to3.eps}
    \end{aligned} \times (-\ell_b + \ell_c + p_3)^2 (\ell_a - \ell_b + \ell_c)^2\\
    =&\frac{0.006944444444}{\epsilon \
^6}-\frac{0.06784533029}{\epsilon ^5}-\frac{0.07453481756}{\epsilon \
^4}+\frac{0.09863437550}{\epsilon
^3}\\
& +\frac{2.930419517}{\epsilon ^2}+\frac{15.33056869}{\epsilon }+74.05486447
\end{aligned}
\end{equation}

\begin{equation}
\begin{aligned}
    I^{\rm NP}_{15}=&\begin{aligned}
        \includegraphics[width=0.23\linewidth]{figure/phi2NplGamma10to3.eps}
    \end{aligned} \times (\ell_a + \ell_c - p_2)^2 (\ell_a - \ell_b + l_c)^2\\
   = &-\frac{0.01157407407}{\epsilon \
^5}+\frac{0.05284838872}{\epsilon ^4}-\frac{0.06050799157}{\epsilon \
^3}-\frac{0.3737181441}{\epsilon
^2}\\
&-\frac{5.886476988}{\epsilon }-35.82086314
\end{aligned}
\end{equation}

\begin{equation}
\begin{aligned}
    I^{\rm NP}_{16}=&\begin{aligned}
        \includegraphics[width=0.23\linewidth]{figure/phi2NplGamma10to3.eps}
    \end{aligned} \times  (\ell_a - \ell_b + \ell_c)^2 \left(-(p_2 + p_3 - \ell_b)^2 + \ell_b^2\right)\\
    =&+\frac{0.009799382716}{\epsilon \
^6}-\frac{0.009573070090}{\epsilon ^5}+\frac{0.3997995824}{\epsilon \
^4}-\frac{2.556473376}{\epsilon
^3}\\
&-\frac{12.04038966}{\epsilon ^2}-\frac{30.74603944}{\epsilon } -639.7868476
\end{aligned}
\end{equation}

\begin{equation}
\begin{aligned}
    I^{\rm NP}_{17}=&\begin{aligned}
    \includegraphics[width=0.23\linewidth]{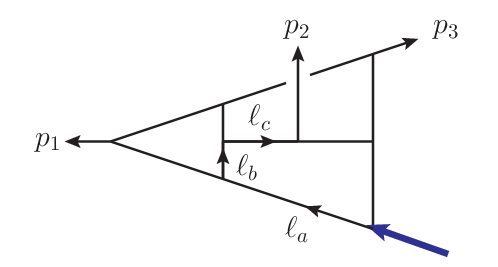}
    \end{aligned} \times  (\ell_a-p_1)^2(\ell_a-p_1-p_2)^2\\
    =&-\frac{0.002314814815}{\epsilon ^6}+\frac{0.05440366012}{\
\epsilon ^5}+\frac{0.2121506031}{\epsilon \
^4}-\frac{1.952521234}{\epsilon
^3}\\
&-\frac{21.14121669}{\epsilon ^2}-\frac{151.1615815}{\epsilon }-948.6049804
\end{aligned}
\end{equation}

\begin{equation}
\begin{aligned}
    I^{\rm NP}_{18}=&\begin{aligned}
    \includegraphics[width=0.23\linewidth]{figure/phi2NplGamma10to4.eps}
    \end{aligned} \times  \left((\ell_a-p_1)^2\right)^2\\
    =&+\frac{0.02314814815}{\epsilon \
^6}-\frac{0.07102297689}{\epsilon ^5}-\frac{0.5148637417}{\epsilon \
^4}-\frac{1.577775234}{\epsilon ^3}\\
&+\frac{2.080912452}{\epsilon
^2}+\frac{40.25823151}{\epsilon }+304.4094179
\end{aligned}
\end{equation}

\begin{equation}
\begin{aligned}
    I^{\rm NP}_{19}=&\begin{aligned}
    \includegraphics[width=0.23\linewidth]{figure/phi2NplGamma10to4.eps}
    \end{aligned} \times  (\ell_a-\ell_c)^2(\ell_a-p_1-p_2)^2\\
   = &-\frac{0.04513888889}{\epsilon \
^6}-\frac{0.02057509937}{\epsilon ^5}+\frac{0.8185749116}{\epsilon \
^4}+\frac{5.008700101}{\epsilon ^3}\\
&+\frac{27.48344200}{\epsilon
^2}+\frac{157.7477751}{\epsilon }+924.6466001
\end{aligned}
\end{equation}

\begin{equation}
\begin{aligned}
    I^{\rm NP}_{20}=&\begin{aligned}
    \includegraphics[width=0.23\linewidth]{figure/phi2NplGamma10to4.eps}
    \end{aligned} \times  (\ell_a-\ell_c)^2(\ell_a-p_1)^2\\
    =&+\frac{0.02314814815}{\epsilon \
^6}-\frac{0.09391073290}{\epsilon ^5}-\frac{0.5453892304}{\epsilon \
^4}-\frac{0.7360772736}{\epsilon ^3}\\
&+\frac{9.531264206}{\epsilon
^2}+\frac{90.70451489}{\epsilon }+581.2346817
\end{aligned}
\end{equation}

\begin{equation}
\begin{aligned}
    I^{\rm NP}_{21}=&\begin{aligned}
    \includegraphics[width=0.23\linewidth]{figure/phi2NplGamma10to4.eps}
    \end{aligned} \times  (\ell_a-p_1)^2\\
    =&-\frac{0.03055555556}{\epsilon \
^6}-\frac{0.01275402850}{\epsilon ^5}+\frac{0.006576522622}{\epsilon \
^4}+\frac{0.03186263329}{\epsilon
^3}\\
&+\frac{1.081865979}{\epsilon ^2}+\frac{20.79582622}{\epsilon }+217.3264489
\end{aligned}
\end{equation}

\begin{equation}
\begin{aligned}
    I^{\rm NP}_{22}=&\begin{aligned}
    \includegraphics[width=0.23\linewidth]{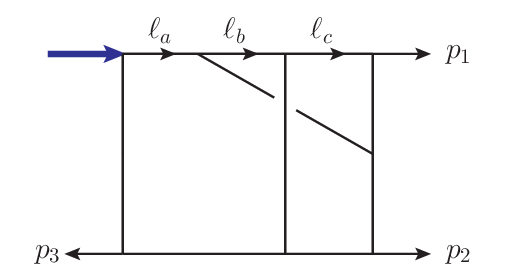}
    \end{aligned} \times  \left((\ell_a-p_1)^2\right)^2\\
   = &+\frac{0.006944444444}{\epsilon \
^6}-\frac{0.06246579507}{\epsilon ^5}-\frac{0.2813601462}{\epsilon \
^4}+\frac{0.2022460374}{\epsilon
^3}\\
&+\frac{15.39827437}{\epsilon ^2}+\frac{143.5286351}{\epsilon }+1046.079927
\end{aligned}
\end{equation}

\begin{equation}
\begin{aligned}
    I^{\rm NP}_{23}=&\begin{aligned}
    \includegraphics[width=0.23\linewidth]{figure/phi2NplGamma10to5.eps}
    \end{aligned} \times  (\ell_a-p_1)^2\\
   = &-\frac{0.008912037037}{\epsilon \
^6}-\frac{0.03705732274}{\epsilon ^5}-\frac{0.2699836198}{\epsilon \
^4}+\frac{0.6353113302}{\epsilon
^3}\\
&+\frac{15.82543009}{\epsilon ^2}+\frac{136.4660569}{\epsilon }+950.1158561
\end{aligned}
\end{equation}

\begin{equation}
\begin{aligned}
    I^{\rm NP}_{24}=&\begin{aligned}
    \includegraphics[width=0.23\linewidth]{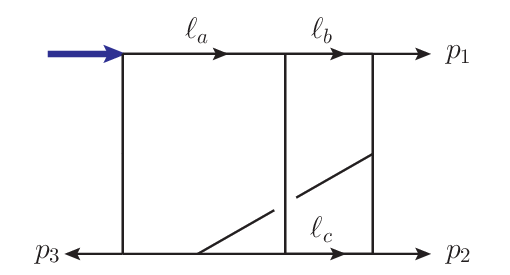}
    \end{aligned} \times  (\ell_a-\ell_b-p_2)^2\\
    =&-\frac{0.01006944444}{\epsilon \
^6}-\frac{0.02254791535}{\epsilon ^5}-\frac{0.3642224231}{\epsilon \
^4}-\frac{1.940390971}{\epsilon ^3} \\
& -\frac{8.326781592}{\epsilon
^2}-\frac{32.88602634}{\epsilon }-142.5708645
\end{aligned}
\end{equation}

\begin{equation}
\begin{aligned}
    I^{\rm NP}_{25}=&\begin{aligned}
    \includegraphics[width=0.23\linewidth]{figure/phi2NplGamma10to6.eps}
    \end{aligned} \times  (s_{12}- \ell_a^2)(-\ell_b + p_1 + p_2)^2\\
    =&\frac{0.03888888889}{\epsilon \
^6}-\frac{0.03814335984}{\epsilon ^5}-\frac{0.9592615659}{\epsilon \
^4}-\frac{5.665699033}{\epsilon ^3}\\
& -\frac{35.28990092}{\epsilon
^2}-\frac{243.2021442}{\epsilon }-1673.609071
\end{aligned}
\end{equation}

\begin{equation}
\begin{aligned}
    I^{\rm NP}_{26}=&\begin{aligned}
    \includegraphics[width=0.23\linewidth]{figure/phi2NplGamma10to6.eps}
    \end{aligned} \times  (\ell_a - p_1)^2 \left((\ell_a - p_1)^2 - (\ell_b - p_1)^2 - (\ell_a - p_1 - p_2)^2 - (\ell_a - \ell_b)^2\right)\\
    =&\frac{0.009027777778}{\epsilon \
^6}+\frac{0.004115019874}{\epsilon ^5}-\frac{0.3955662066}{\epsilon \
^4}-\frac{1.240635884}{\epsilon
^3}\\
& -\frac{7.577039255}{\epsilon ^2}-\frac{63.27334604}{\epsilon }-518.9849090
\end{aligned}
\end{equation}


\begin{equation}
\begin{aligned}
    I^{\rm NP}_{27}=&\begin{aligned}
    \includegraphics[width=0.23\linewidth]{figure/phi2NplGamma10to6.eps}
    \end{aligned} \times  (\ell_a - p_1)^2  (\ell_b - p_1)^2 \\
    =&-\frac{0.006944444444}{\epsilon \
^5}+\frac{0.02627503247}{\epsilon ^4}+\frac{0.3956079668}{\epsilon \
^3}+\frac{3.317841000}{\epsilon ^2} \\
    &+\frac{19.73712993}{\epsilon
}+112.6684764
\end{aligned}
\end{equation}

\begin{equation}
\begin{aligned}
    I^{\rm NP}_{28}=&\begin{aligned}
    \includegraphics[width=0.23\linewidth]{figure/phi2NplGamma10to6.eps}
    \end{aligned} \times  \ell_a^2  (\ell_b - p_1-p_2)^2 \\
    =&\frac{0.1243055556}{\epsilon \
^5}-\frac{0.4419056126}{\epsilon ^4}+\frac{6.383061420}{\epsilon ^3} -\frac{20.25286847}{\epsilon ^2} \\
    &+\frac{195.9136523}{\epsilon
}-867.0739223
\end{aligned}
\end{equation}

\begin{equation}\label{eq:NPzero1}
\begin{aligned}
    I^{\rm NP}_{29}=&\begin{aligned}
    \includegraphics[width=0.23\linewidth]{figure/phi2NplGamma10to6.eps}
    \end{aligned} \times  (\ell_a - \ell_b)^2 \\
    =&-\frac{0.04722222222}{\epsilon \
^5}+\frac{0.04917232108}{\epsilon ^4}+\frac{1.239906872}{\epsilon \
^3} -\frac{8.752864003}{\epsilon ^2} \\
    &+\frac{129.6066269}{\epsilon
}-586.5917818
\end{aligned}
\end{equation}

\begin{equation}
\begin{aligned}
    I^{\rm NP}_{30}=&\begin{aligned}
    \includegraphics[width=0.23\linewidth]{figure/phi2NplGamma10to6.eps}
    \end{aligned} \times    (\ell_b - p_1)^2 \\
=&-\frac{0.005208333333}{\epsilon ^6}-\frac{0.007557897755}{\
\epsilon ^5}+\frac{0.03670809281}{\epsilon \
^4}+\frac{1.198027500}{\epsilon
^3}\\
& +\frac{10.96058376}{\epsilon ^2}+\frac{79.76107702}{\epsilon }+518.6921866
\end{aligned}
\end{equation}

\begin{equation}
\begin{aligned}
    I^{\rm NP}_{31}=&\begin{aligned}
    \includegraphics[width=0.23\linewidth]{figure/phi2NplGamma10to6.eps}
    \end{aligned} \times  (\ell_a-p_1-p_2)^2(\ell_a-\ell_b)^2\\
    =&-\frac{0.006866326804}{\epsilon ^5}+\frac{0.08988715904}{\
\epsilon ^4}-\frac{0.3949600076}{\epsilon \
^3}-\frac{2.338326746}{\epsilon^2} \\
    &-\frac{10.33275643}{\epsilon } -32.51969920
\end{aligned}
\end{equation}

\begin{equation}
\begin{aligned}
    I^{\rm NP}_{32}=&\begin{aligned}
    \includegraphics[width=0.23\linewidth]{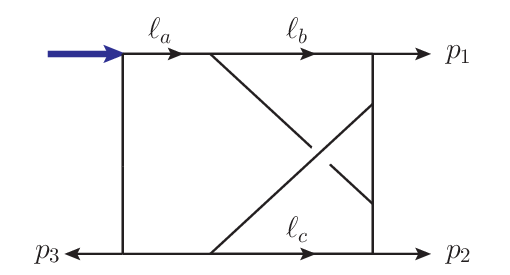}
    \end{aligned} \times  \left((\ell_a-p_1)^2\right)^2\\
   = &+\frac{0.02430555556}{\epsilon \
^6}+\frac{0.06389679815}{\epsilon ^5}+\frac{0.04416397528}{\epsilon \
^4}+\frac{0.6965715244}{\epsilon
^3}\\
&+\frac{12.84868788}{\epsilon ^2}+\frac{106.0216047}{\epsilon }+685.0980200
\end{aligned}
\end{equation}

\begin{equation}
\begin{aligned}
    I^{\rm NP}_{33}=&\begin{aligned}
    \includegraphics[width=0.23\linewidth]{figure/phi2NplGamma10to7.eps}
    \end{aligned} \times  (\ell_a-p_1)^2\\
    =&+\frac{0.02546296296}{\epsilon \
^6}-\frac{0.008291992672}{\epsilon ^5}+\frac{0.2706584397}{\epsilon \
^4}-\frac{1.955326531}{\epsilon ^3}\\
&+\frac{25.48473558}{\epsilon
^2}+\frac{11.43562088}{\epsilon }+1079.405340
\end{aligned}
\end{equation}

\begin{equation}
\begin{aligned}
    I^{\rm NP}_{34}=&\begin{aligned}
        \includegraphics[width=0.23\linewidth]{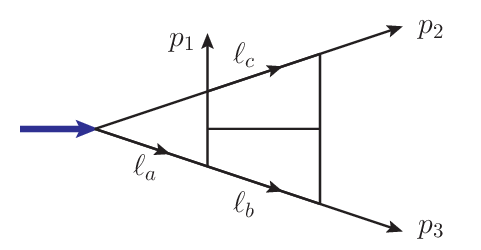}
    \end{aligned}\times (\ell_a-p_3)^2\\
    =&-\frac{0.01144387801}{\epsilon \
^5}+\frac{0.07737511477}{\epsilon ^4}-\frac{0.06541086481}{\epsilon \
^3}-\frac{0.1908458531}{\epsilon
^2}\\
&-\frac{2.487935703}{\epsilon }-11.78273194
\end{aligned}
\end{equation}

\begin{equation}
\begin{aligned}
    I^{\rm NP}_{35}=&\begin{aligned}
        \includegraphics[width=0.23\linewidth]{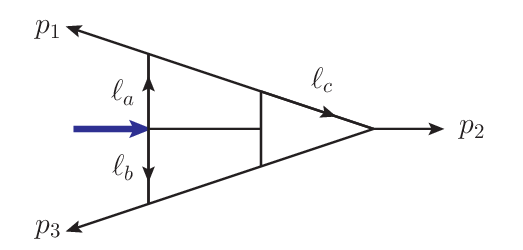}
    \end{aligned}\times (\ell_b-p_2-p_3)^2\\
    =&+\frac{0.04861111111}{\epsilon \
^6}-\frac{0.1010839638}{\epsilon ^5}-\frac{0.3425098100}{\epsilon \
^4}-\frac{0.9284540918}{\epsilon ^3}\\
&-\frac{0.7016321430}{\epsilon
^2}-\frac{2.161180083}{\epsilon }-14.51695239
\end{aligned}
\end{equation}

\begin{equation}
\begin{aligned}
    I^{\rm NP}_{36}=&\begin{aligned}
        \includegraphics[width=0.23\linewidth]{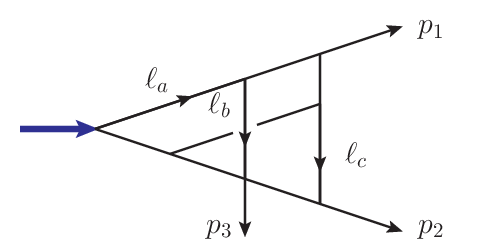}
    \end{aligned}\times \\
    =&+\frac{0.01273148148}{\epsilon \
^5}-\frac{0.06226964468}{\epsilon ^4}-\frac{0.08095467186}{\epsilon \
^3}+\frac{0.2313400069}{\epsilon^2} \\
    &+\frac{0.7762286087}{\epsilon }-3.806749978
\end{aligned}
\end{equation}

\begin{equation}
\begin{aligned}
    I^{\rm NP}_{37}=&\begin{aligned}
        \includegraphics[width=0.23\linewidth]{figure/phi2NplGamma9to3.eps}
    \end{aligned}\times (\ell_b-p_3)^2 \\
    =& -\frac{0.004499433563}{\epsilon^5}+\frac{0.05252605425}{\epsilon^4}-\frac{0.1162423852}{\epsilon^3}-\frac{0.06526801051}{\epsilon^2} \\
    &-\frac{0.3929346129}{\epsilon }+1.779113208
\end{aligned}
\end{equation}

\begin{equation}
\begin{aligned}
    I^{\rm NP}_{38}=&\begin{aligned}
        \includegraphics[width=0.23\linewidth]{figure/phi2NplGamma9to3.eps}
    \end{aligned}\times (\ell_b-p_2-p_3)^2 \\
    =&-\frac{0.02288775601}{\epsilon \
^5}+\frac{0.1287104831}{\epsilon ^4}+\frac{0.05003666587}{\epsilon \
^3}-\frac{0.2998697024}{\epsilon
^2} \\
    &-\frac{3.187512584}{\epsilon } -7.913302690
\end{aligned}
\end{equation}

\begin{equation}
\begin{aligned}
    I^{\rm NP}_{39}=&\begin{aligned}
        \includegraphics[width=0.23\linewidth]{figure/phi2NplGamma9to3.eps}
    \end{aligned}\times (\ell_a-\ell_b-p_1-p_2)^2 \\
    =& +\frac{0.03771715503}{\epsilon \
^4}-\frac{0.1223564048}{\epsilon ^3}-\frac{0.3750614030}{\epsilon^2}+\frac{1.211044132}{\epsilon } \\
    &+19.60355026
\end{aligned}
\end{equation}

\begin{equation}
\begin{aligned}
    I^{\rm NP}_{40}=&\begin{aligned}
        \includegraphics[width=0.23\linewidth]{figure/phi2NplGamma9to3.eps}
    \end{aligned}\times \left( (\ell_a-p_1)^2 - (\ell_a-p_3)^2 \right)\\
    =& +\frac{0.009389455326}{\epsilon \
^5}-\frac{0.02270917538}{\epsilon ^4}-\frac{0.1739747206}{\epsilon \
^3}-\frac{0.1340198906}{\epsilon
^2} \\
    &+\frac{3.567719152}{\epsilon }+28.50509928
\end{aligned}
\end{equation}

\begin{equation}
\begin{aligned}
    I^{\rm NP}_{41}=&\begin{aligned}
        \includegraphics[width=0.23\linewidth]{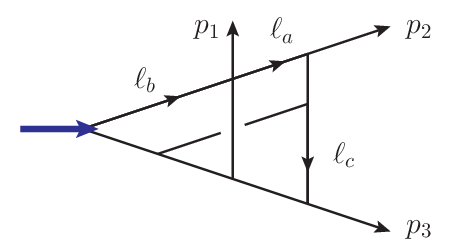}
    \end{aligned}\times (\ell_a-\ell_b+p_1+p_3)^2\\
    =&-\frac{0.01144387801}{\epsilon \
^5}+\frac{0.1025198848}{\epsilon ^4}-\frac{0.1469818013}{\epsilon \
^3}-\frac{0.5839314324}{\epsilon ^2}\\
&-\frac{1.673263147}{\epsilon
}+3.944288152
\end{aligned}
\end{equation}

\begin{equation}
\begin{aligned}
    I^{\rm NP}_{42}=&\begin{aligned}
        \includegraphics[width=0.23\linewidth]{figure/phi2NplGamma9to4.eps}
    \end{aligned}\times (\ell_a-p_2-p_3)^2\\
    =&+\frac{0.002314814815}{\epsilon ^6}-\frac{0.02007202610}{\
\epsilon ^5}+\frac{0.06622380851}{\epsilon \
^4}+\frac{0.02390993104}{\epsilon
^3}\\
&-\frac{1.116645221}{\epsilon ^2}-\frac{8.391534860}{\epsilon }-34.83743524
\end{aligned}
\end{equation}

\begin{equation}
\begin{aligned}
    I^{\rm NP}_{43}=&\begin{aligned}
        \includegraphics[width=0.23\linewidth]{figure/phi2NplGamma9to5.eps}
    \end{aligned}\times (\ell_a-p_2-p_3)^2\\
    =&+\frac{0.1562500000}{\epsilon \
^6}-\frac{0.2333617168}{\epsilon ^5}-\frac{1.420241193}{\epsilon ^4}-\
\frac{1.957240838}{\epsilon ^3}\\
&+\frac{8.026361751}{\epsilon
^2}+\frac{44.51461807}{\epsilon }+119.2673216
\end{aligned}
\end{equation}

\begin{equation}
\begin{aligned}
    I^{\rm NP}_{44}=&\begin{aligned}
        \includegraphics[width=0.23\linewidth]{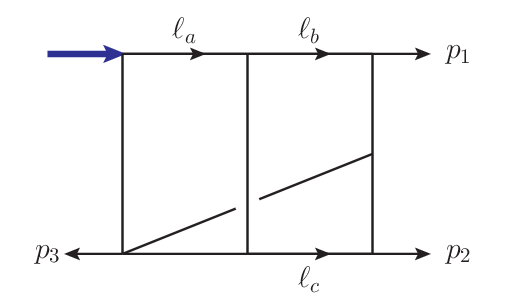}
    \end{aligned}\times (\ell_a-p_1-p_2)^2\\
    =&+\frac{0.006944444444}{\epsilon \
^6}-\frac{0.06784533029}{\epsilon ^5}-\frac{0.07453481756}{\epsilon \
^4}+\frac{0.09863437550}{\epsilon
^3}\\
&+\frac{2.930419517}{\epsilon ^2}+\frac{15.33056869}{\epsilon }+74.05486447
\end{aligned}
\end{equation}

\begin{equation}
\begin{aligned}
    I^{\rm NP}_{45}=&\begin{aligned}
        \includegraphics[width=0.23\linewidth]{figure/phi2NplGamma9to6.eps}
    \end{aligned}\times (\ell_a-p_1)^2\\
    =&+\frac{0.004577551203}{\epsilon \
^5}+\frac{0.03855179072}{\epsilon ^4}-\frac{0.8740631543}{\epsilon \
^3}-\frac{3.924277491}{\epsilon ^2}\\
&-\frac{10.02096613}{\epsilon
}+6.561674883
\end{aligned}
\end{equation}

\begin{equation}
\begin{aligned}
    I^{\rm NP}_{46}=&\begin{aligned}
        \includegraphics[width=0.23\linewidth]{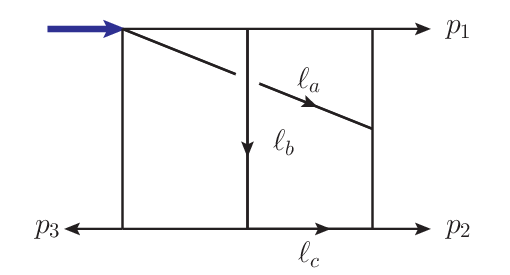}
    \end{aligned}\times (\ell_a-\ell_b+\ell_c-p_2)^2\\
    =&-\frac{0.04305555556}{\epsilon \
^6}+\frac{0.02047765458}{\epsilon ^5}+\frac{1.079792439}{\epsilon \
^4}+\frac{1.378399612}{\epsilon ^3}\\
&-\frac{6.904008819}{\epsilon
^2}-\frac{58.97507398}{\epsilon }-293.5418836
\end{aligned}
\end{equation}

\begin{equation}
\begin{aligned}
    I^{\rm NP}_{47}=&\begin{aligned}
        \includegraphics[width=0.23\linewidth]{figure/phi2NplGamma9to7.eps}
    \end{aligned}\times \ell_a\cdot \ell_b\\
    =&+\frac{0.02986111111}{\epsilon \
^6}-\frac{0.06209443493}{\epsilon ^5}-\frac{0.4276876808}{\epsilon \
^4}-\frac{0.06863992757}{\epsilon
^3}\\
&+\frac{7.287445633}{\epsilon ^2}+\frac{52.86752669}{\epsilon }+311.0917928
\end{aligned}
\end{equation}

\begin{equation}
\begin{aligned}
    I^{\rm NP}_{48}=&\begin{aligned}
        \includegraphics[width=0.23\linewidth]{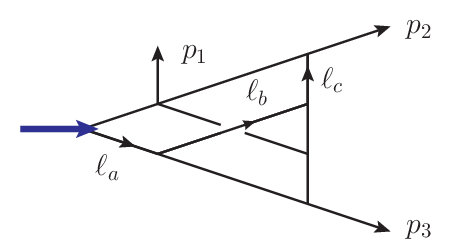}
    \end{aligned}\times (\ell_a - p_3)^2\\
    =&-\frac{0.002314814815}{\epsilon ^6}+\frac{0.02007202610}{\
\epsilon ^5}+\frac{0.08030190972}{\epsilon \
^4}-\frac{0.7266717980}{\epsilon
^3}\\
&-\frac{6.269861622}{\epsilon ^2}-\frac{33.54858950}{\epsilon }-152.3561653
\end{aligned}
\end{equation}


\begin{equation}
\begin{aligned}
    I^{\rm NP}_{49}=&\begin{aligned}
        \includegraphics[width=0.23\linewidth]{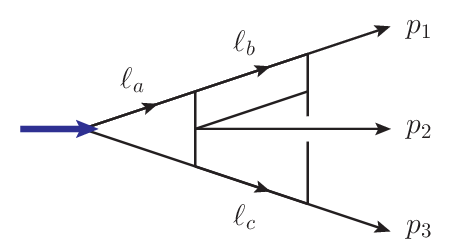}
    \end{aligned} \\
    =&-\frac{0.009027777778}{\epsilon ^6}-\frac{0.02700277589}{\
\epsilon ^5}-\frac{0.3096826614}{\epsilon^4}-\frac{1.497430653}{\epsilon^3}\\
&-\frac{6.191416256}{\epsilon ^2}-\frac{19.94394828}{\epsilon }-14.39206650
\end{aligned}
\end{equation}

\begin{equation}
\begin{aligned}
    I^{\rm NP}_{50}=&\begin{aligned}
        \includegraphics[width=0.23\linewidth]{figure/phi2NplGamma9to10.eps}
    \end{aligned}\times (\ell_a-p_1)^2 \\
    =&-\frac{0.02288775601}{\epsilon ^5}+\frac{0.001196680142}{\
\epsilon ^4}-\frac{0.3754600853}{\epsilon \
^3}-\frac{1.643869853}{\epsilon
^2}\\
&-\frac{7.567496564}{\epsilon }-13.07725291
\end{aligned}
\end{equation}

\begin{equation}
\begin{aligned}
    I^{\rm NP}_{51}=&\begin{aligned}
        \includegraphics[width=0.23\linewidth]{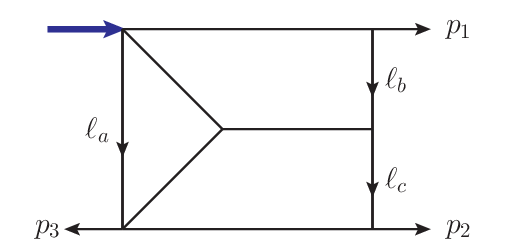}
    \end{aligned}\\
    =& -\frac{0.03472222222}{\epsilon \
^5}+\frac{0.1424284522}{\epsilon ^4}+\frac{0.02394649346}{\epsilon \
^3}-\frac{0.4791070732}{\epsilon
^2}\\
&-\frac{3.819407715}{\epsilon }-10.33009288
\end{aligned}
\end{equation}

\begin{equation}
\begin{aligned}
    I^{\rm NP}_{52}=&\begin{aligned}
        \includegraphics[width=0.23\linewidth]{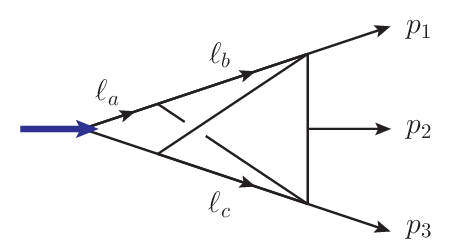}
    \end{aligned}\\
    =&  -\frac{1.770044742}{\epsilon }-11.76750928
\end{aligned}
\end{equation}

\begin{equation}
\begin{aligned}
    I^{\rm NP}_{53}=&\begin{aligned}
        \includegraphics[width=0.23\linewidth]{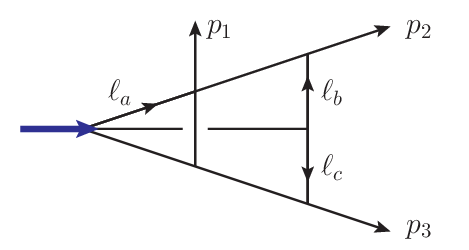}
    \end{aligned}\\
    =&-\frac{0.02288775601}{\epsilon \
^5}+\frac{0.1790000231}{\epsilon ^4}-\frac{0.1131052072}{\epsilon \
^3}-\frac{1.474692214}{\epsilon ^2}\\
&-\frac{5.751773424}{\epsilon
}-3.514390529
\end{aligned}
\end{equation}

\begin{equation}
\begin{aligned}
    I^{\rm NP}_{54}=&\begin{aligned}
        \includegraphics[width=0.23\linewidth]{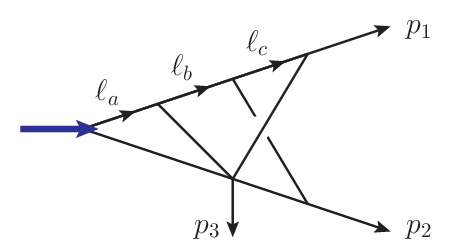}
    \end{aligned}\\
    =& -\frac{0.01157407407}{\epsilon \
^5}+\frac{0.07799315874}{\epsilon ^4}-\frac{0.1034125732}{\epsilon \
^3}-\frac{0.2442150034}{\epsilon ^2}\\
&-\frac{1.091330793}{\epsilon
}+1.394464631
\end{aligned}
\end{equation}

\begin{equation}
\begin{aligned}
    I^{\rm NP}_{55}=&\begin{aligned}
        \includegraphics[width=0.23\linewidth]{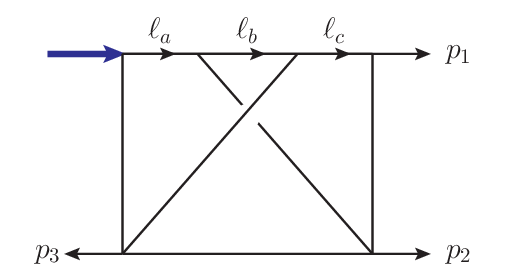}
    \end{aligned}\\
    =& -\frac{0.03866635486}{\epsilon \
^3}+\frac{0.2951965652}{\epsilon ^2}+\frac{1.798905340}{\epsilon }+9.003276865
\end{aligned}
\end{equation}

\begin{equation}
\begin{aligned}
    I^{\rm NP}_{56}=&\begin{aligned}
        \includegraphics[width=0.23\linewidth]{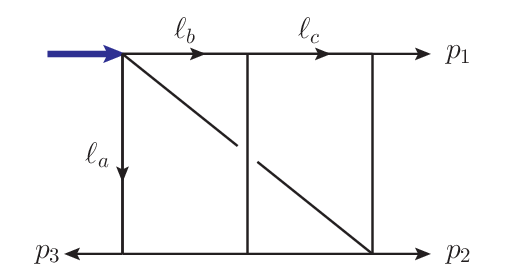}
    \end{aligned}\\
    =& +\frac{0.01041666667}{\epsilon ^6}+\frac{0.001226906621}{\
\epsilon ^5}-\frac{0.02086612068}{\epsilon \
^4}-\frac{1.450173648}{\epsilon
^3}\\
&-\frac{13.94768363}{\epsilon ^2}-\frac{108.0193124}{\epsilon }-726.4945266
\end{aligned}
\end{equation}

\begin{equation}
\begin{aligned}
    I^{\rm NP}_{57}=&\begin{aligned}
        \includegraphics[width=0.23\linewidth]{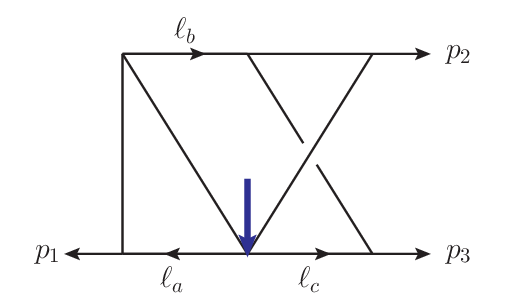}
    \end{aligned}\\
    =& +\frac{0.1666666667}{\epsilon \
^5}-\frac{0.6799069236}{\epsilon ^4}-\frac{0.4799078429}{\epsilon \
^3}+\frac{10.50986209}{\epsilon ^2}\\
&-\frac{151.1442636}{\epsilon
}+1266.966539
\end{aligned}
\end{equation}

\begin{equation}
\begin{aligned}
    I^{\rm NP}_{58}=&\begin{aligned}
        \includegraphics[width=0.23\linewidth]{figure/phi2NplGamma10to4.eps}
    \end{aligned} \times  (\ell_a-\ell_c)^2 (-s_{23}+\left(\ell_a-p_1-p_2-p_3\right)^2 ) \\
    =& - \frac{0.09884259259}{\epsilon^6} - \frac{0.0004524110009}{\epsilon^5} + \frac{1.452766570}{\epsilon^4} + \frac{7.291529822}{\epsilon^3} \\
     & + \frac{36.318452263370}{\epsilon^2} + \frac{227.0170637}{\epsilon} + 1508.502038 
\end{aligned}
\end{equation}

\begin{equation}\label{eq:NPzero2}
\begin{aligned}
    I^{\rm NP}_{59}=&\begin{aligned}
        \includegraphics[width=0.23\linewidth]{figure/phi2NplGamma10to3.eps}
    \end{aligned} \times ( \ell_c -p_1-p_2)^2  \\
    =& - \frac{0.06222993827}{\epsilon^6} - \frac{0.1714450039}{\epsilon^5} + \frac{1.819541239}{\epsilon^4} + \frac{5.751038831}{\epsilon^3} \\
     & + \frac{5.525406546}{\epsilon^2} + \frac{29.97585133}{\epsilon} -1292.048160
\end{aligned}
\end{equation}

We comment that for two of the above non-planar integral basis (\eqref{eq:NPzero1} and \eqref{eq:NPzero2}), their coefficients $c_j$ in the form factor \eqref{eq:simpInt2np} are 
\begin{equation}
c_{29}={1\over4} s_{12}^2(s_{23}-s_{13}) \,, \qquad c_{59}=-{1\over2}s_{12}s_{23}(s_{12}-s_{23}) \,,
\end{equation} 
both of which are zero for the special kinematics where $s_{12}=s_{23}=s_{13}$. Therefore, these two integrals do not contribute to the form factor at this specific kinematic point but will contribute for more general kinematics.


\begin{thebibliography}{10}

\bibitem{Lin:2021kht}
G.~Lin, G.~Yang, and S.~Zhang, {\it {Three-loop color-kinematics duality:
  24-dimensional solution space induced by new generalized gauge
  transformations}},  \href{http://arxiv.org/abs/2106.05280}{{\tt
  arXiv:2106.05280}}.

\bibitem{Lin:2021qol}
G.~Lin, G.~Yang, and S.~Zhang, {\it {Full-color three-loop three-point form
  factors in \ensuremath{\mathscr{N}} = 4 SYM}},  {\em JHEP} {\bf 03} (2022)
  061, [\href{http://arxiv.org/abs/2111.03021}{{\tt arXiv:2111.03021}}].

\bibitem{Bern:2008qj}
Z.~Bern, J.~J.~M. Carrasco, and H.~Johansson, {\it {New Relations for
  Gauge-Theory Amplitudes}},  {\em Phys. Rev.} {\bf D78} (2008) 085011,
  [\href{http://arxiv.org/abs/0805.3993}{{\tt arXiv:0805.3993}}].

\bibitem{Bern:2010ue}
Z.~Bern, J.~J.~M. Carrasco, and H.~Johansson, {\it {Perturbative Quantum
  Gravity as a Double Copy of Gauge Theory}},  {\em Phys.Rev.Lett.} {\bf 105}
  (2010) 061602, [\href{http://arxiv.org/abs/1004.0476}{{\tt
  arXiv:1004.0476}}].

\bibitem{Bern:1994zx}
Z.~Bern, L.~J. Dixon, D.~C. Dunbar, and D.~A. Kosower, {\it {One loop n point
  gauge theory amplitudes, unitarity and collinear limits}},  {\em Nucl.Phys.}
  {\bf B425} (1994) 217--260, [\href{http://arxiv.org/abs/hep-ph/9403226}{{\tt
  hep-ph/9403226}}].

\bibitem{Bern:1994cg}
Z.~Bern, L.~J. Dixon, D.~C. Dunbar, and D.~A. Kosower, {\it {Fusing gauge
  theory tree amplitudes into loop amplitudes}},  {\em Nucl. Phys. B} {\bf 435}
  (1995) 59--101, [\href{http://arxiv.org/abs/hep-ph/9409265}{{\tt
  hep-ph/9409265}}].

\bibitem{Britto:2004nc}
R.~Britto, F.~Cachazo, and B.~Feng, {\it {Generalized unitarity and one-loop
  amplitudes in N=4 super-Yang-Mills}},  {\em Nucl.Phys.} {\bf B725} (2005)
  275--305, [\href{http://arxiv.org/abs/hep-th/0412103}{{\tt hep-th/0412103}}].

\bibitem{DiVita:2014pza}
S.~Di~Vita, P.~Mastrolia, U.~Schubert, and V.~Yundin, {\it {Three-loop master
  integrals for ladder-box diagrams with one massive leg}},  {\em JHEP} {\bf
  09} (2014) 148, [\href{http://arxiv.org/abs/1408.3107}{{\tt
  arXiv:1408.3107}}].

\bibitem{Canko:2021xmn}
D.~D. Canko and N.~Syrrakos, {\it {Planar three-loop master integrals for 2
  \textrightarrow{} 2 processes with one external massive particle}},  {\em
  JHEP} {\bf 04} (2022) 134, [\href{http://arxiv.org/abs/2112.14275}{{\tt
  arXiv:2112.14275}}].

\bibitem{Henn:2023vbd}
J.~M. Henn, J.~Lim, and W.~J. Torres~Bobadilla, {\it {First look at the
  evaluation of three-loop non-planar Feynman diagrams for Higgs plus jet
  production}},  {\em JHEP} {\bf 05} (2023) 026,
  [\href{http://arxiv.org/abs/2302.12776}{{\tt arXiv:2302.12776}}].

\bibitem{Smirnov:2015mct}
A.~V. Smirnov, {\it {FIESTA4: Optimized Feynman integral calculations with GPU
  support}},  {\em Comput. Phys. Commun.} {\bf 204} (2016) 189--199,
  [\href{http://arxiv.org/abs/1511.03614}{{\tt arXiv:1511.03614}}].

\bibitem{Borowka:2017idc}
S.~Borowka, G.~Heinrich, S.~Jahn, S.~P. Jones, M.~Kerner, J.~Schlenk, and
  T.~Zirke, {\it {pySecDec: a toolbox for the numerical evaluation of
  multi-scale integrals}},  {\em Comput. Phys. Commun.} {\bf 222} (2018)
  313--326, [\href{http://arxiv.org/abs/1703.09692}{{\tt arXiv:1703.09692}}].

\bibitem{Liu:2017jxz}
X.~Liu, Y.-Q. Ma, and C.-Y. Wang, {\it {A Systematic and Efficient Method to
  Compute Multi-loop Master Integrals}},  {\em Phys. Lett. B} {\bf 779} (2018)
  353--357, [\href{http://arxiv.org/abs/1711.09572}{{\tt arXiv:1711.09572}}].

\bibitem{Liu:2021wks}
X.~Liu and Y.-Q. Ma, {\it {Multiloop corrections for collider processes using
  auxiliary mass flow}},  \href{http://arxiv.org/abs/2107.01864}{{\tt
  arXiv:2107.01864}}.

\bibitem{Liu:2022mfb}
Z.-F. Liu and Y.-Q. Ma, {\it {Determining Feynman Integrals with Only Input
  from Linear Algebra}},  {\em Phys. Rev. Lett.} {\bf 129} (2022), no.~22
  222001, [\href{http://arxiv.org/abs/2201.11637}{{\tt arXiv:2201.11637}}].

\bibitem{Liu:2022chg}
X.~Liu and Y.-Q. Ma, {\it {AMFlow: A Mathematica package for Feynman integrals
  computation via auxiliary mass flow}},  {\em Comput. Phys. Commun.} {\bf 283}
  (2023) 108565, [\href{http://arxiv.org/abs/2201.11669}{{\tt
  arXiv:2201.11669}}].

\bibitem{Dixon:2020bbt}
L.~J. Dixon, A.~J. McLeod, and M.~Wilhelm, {\it {A Three-Point Form Factor
  Through Five Loops}},  {\em JHEP} {\bf 04} (2021) 147,
  [\href{http://arxiv.org/abs/2012.12286}{{\tt arXiv:2012.12286}}].

\bibitem{Bern:2005iz}
Z.~Bern, L.~J. Dixon, and V.~A. Smirnov, {\it {Iteration of planar amplitudes
  in maximally supersymmetric Yang-Mills theory at three loops and beyond}},
  {\em Phys. Rev.} {\bf D72} (2005) 085001,
  [\href{http://arxiv.org/abs/hep-th/0505205}{{\tt hep-th/0505205}}].

\bibitem{Spradlin:2008uu}
M.~Spradlin, A.~Volovich, and C.~Wen, {\it {Three-Loop Leading Singularities
  and BDS Ansatz for Five Particles}},  {\em Phys. Rev.} {\bf D78} (2008)
  085025, [\href{http://arxiv.org/abs/0808.1054}{{\tt arXiv:0808.1054}}].

\bibitem{Eden:2011yp}
B.~Eden, P.~Heslop, G.~P. Korchemsky, and E.~Sokatchev, {\it {The
  super-correlator/super-amplitude duality: Part I}},  {\em Nucl. Phys. B} {\bf
  869} (2013) 329--377, [\href{http://arxiv.org/abs/1103.3714}{{\tt
  arXiv:1103.3714}}].

\bibitem{Brandhuber:2011tv}
A.~Brandhuber, O.~Gurdogan, R.~Mooney, G.~Travaglini, and G.~Yang, {\it
  {Harmony of Super Form Factors}},  {\em JHEP} {\bf 1110} (2011) 046,
  [\href{http://arxiv.org/abs/1107.5067}{{\tt arXiv:1107.5067}}].

\bibitem{Liu:2018dmc}
X.~Liu and Y.-Q. Ma, {\it {Determining arbitrary Feynman integrals by vacuum
  integrals}},  {\em Phys. Rev. D} {\bf 99} (2019), no.~7 071501,
  [\href{http://arxiv.org/abs/1801.10523}{{\tt arXiv:1801.10523}}].

\bibitem{Guan:2019bcx}
X.~Guan, X.~Liu, and Y.-Q. Ma, {\it {Complete reduction of integrals in
  two-loop five-light-parton scattering amplitudes}},  {\em Chin. Phys. C} {\bf
  44} (2020), no.~9 093106, [\href{http://arxiv.org/abs/1912.09294}{{\tt
  arXiv:1912.09294}}].

\bibitem{www:Blade}
{\em \url{https://gitlab.com/multiloop-pku/blade}}.

\bibitem{Lin:2021pne}
G.~Lin and G.~Yang, {\it {Double Copy of Form Factors and Higgs Amplitudes: An
  Example of Turning Spurious Poles in Yang-Mills into Physical Poles in
  Gravity}},  \href{http://arxiv.org/abs/2111.12719}{{\tt arXiv:2111.12719}}.

\bibitem{Brandhuber:2010ad}
A.~Brandhuber, B.~Spence, G.~Travaglini, and G.~Yang, {\it {Form Factors in N=4
  Super Yang-Mills and Periodic Wilson Loops}},  {\em JHEP} {\bf 1101} (2011)
  134, [\href{http://arxiv.org/abs/1011.1899}{{\tt arXiv:1011.1899}}].

\bibitem{Brandhuber:2012vm}
A.~Brandhuber, G.~Travaglini, and G.~Yang, {\it {Analytic two-loop form factors
  in N=4 SYM}},  {\em JHEP} {\bf 1205} (2012) 082,
  [\href{http://arxiv.org/abs/1201.4170}{{\tt arXiv:1201.4170}}].

\bibitem{Dixon:2021tdw}
L.~J. Dixon, O.~Gurdogan, A.~J. McLeod, and M.~Wilhelm, {\it {Folding
  Amplitudes into Form Factors: An Antipodal Duality}},  {\em Phys. Rev. Lett.}
  {\bf 128} (2022), no.~11 111602, [\href{http://arxiv.org/abs/2112.06243}{{\tt
  arXiv:2112.06243}}].

\bibitem{Gehrmann:2011xn}
T.~Gehrmann, J.~M. Henn, and T.~Huber, {\it {The three-loop form factor in N=4
  super Yang-Mills}},  {\em JHEP} {\bf 03} (2012) 101,
  [\href{http://arxiv.org/abs/1112.4524}{{\tt arXiv:1112.4524}}].

\bibitem{Almelid:2015jia}
{\O}.~Almelid, C.~Duhr, and E.~Gardi, {\it {Three-loop corrections to the soft
  anomalous dimension in multileg scattering}},  {\em Phys. Rev. Lett.} {\bf
  117} (2016), no.~17 172002, [\href{http://arxiv.org/abs/1507.00047}{{\tt
  arXiv:1507.00047}}].

\bibitem{Ellis:1975ap}
J.~R. Ellis, M.~K. Gaillard, and D.~V. Nanopoulos, {\it {A Phenomenological
  Profile of the Higgs Boson}},  {\em Nucl. Phys.} {\bf B106} (1976) 292.

\bibitem{Georgi:1977gs}
H.~M. Georgi, S.~L. Glashow, M.~E. Machacek, and D.~V. Nanopoulos, {\it {Higgs
  Bosons from Two Gluon Annihilation in Proton Proton Collisions}},  {\em Phys.
  Rev. Lett.} {\bf 40} (1978) 692.

\bibitem{Wilczek:1977zn}
F.~Wilczek, {\it {Decays of Heavy Vector Mesons Into Higgs Particles}},  {\em
  Phys. Rev. Lett.} {\bf 39} (1977) 1304.

\bibitem{Shifman:1979eb}
M.~A. Shifman, A.~I. Vainshtein, M.~B. Voloshin, and V.~I. Zakharov, {\it
  {Low-Energy Theorems for Higgs Boson Couplings to Photons}},  {\em Sov. J.
  Nucl. Phys.} {\bf 30} (1979) 711--716. [Yad. Fiz.30,1368(1979)].

\bibitem{Kniehl:1995tn}
B.~A. Kniehl and M.~Spira, {\it {Low-energy theorems in Higgs physics}},  {\em
  Z. Phys.} {\bf C69} (1995) 77--88,
  [\href{http://arxiv.org/abs/hep-ph/9505225}{{\tt hep-ph/9505225}}].

\bibitem{Gehrmann:2011aa}
T.~Gehrmann, M.~Jaquier, E.~Glover, and A.~Koukoutsakis, {\it {Two-Loop QCD
  Corrections to the Helicity Amplitudes for $H \to$ 3 partons}},  {\em JHEP}
  {\bf 1202} (2012) 056, [\href{http://arxiv.org/abs/1112.3554}{{\tt
  arXiv:1112.3554}}].

\bibitem{Kotikov:2002ab}
A.~V. Kotikov and L.~N. Lipatov, {\it {DGLAP and BFKL equations in the $N=4$
  supersymmetric gauge theory}},  {\em Nucl. Phys.} {\bf B661} (2003) 19--61,
  [\href{http://arxiv.org/abs/hep-ph/0208220}{{\tt hep-ph/0208220}}]. [Erratum:
  Nucl. Phys.B685,405(2004)].

\bibitem{Kotikov:2004er}
A.~Kotikov, L.~Lipatov, A.~Onishchenko, and V.~Velizhanin, {\it {Three loop
  universal anomalous dimension of the Wilson operators in N=4 SUSY Yang-Mills
  model}},  {\em Phys.Lett.} {\bf B595} (2004) 521--529,
  [\href{http://arxiv.org/abs/hep-th/0404092}{{\tt hep-th/0404092}}].

\bibitem{Kotikov:1990kg}
A.~V. Kotikov, {\it {Differential equations method: New technique for massive
  Feynman diagrams calculation}},  {\em Phys. Lett. B} {\bf 254} (1991)
  158--164.

\bibitem{Remiddi:1997ny}
E.~Remiddi, {\it {Differential equations for Feynman graph amplitudes}},  {\em
  Nuovo Cim. A} {\bf 110} (1997) 1435--1452,
  [\href{http://arxiv.org/abs/hep-th/9711188}{{\tt hep-th/9711188}}].

\bibitem{Henn:2013pwa}
J.~M. Henn, {\it {Multiloop integrals in dimensional regularization made
  simple}},  {\em Phys. Rev. Lett.} {\bf 110} (2013) 251601,
  [\href{http://arxiv.org/abs/1304.1806}{{\tt arXiv:1304.1806}}].

\bibitem{Lee:2014ioa}
R.~N. Lee, {\it {Reducing differential equations for multiloop master
  integrals}},  {\em JHEP} {\bf 04} (2015) 108,
  [\href{http://arxiv.org/abs/1411.0911}{{\tt arXiv:1411.0911}}].

\bibitem{Moriello:2019yhu}
F.~Moriello, {\it {Generalised power series expansions for the elliptic planar
  families of Higgs + jet production at two loops}},  {\em JHEP} {\bf 01}
  (2020) 150, [\href{http://arxiv.org/abs/1907.13234}{{\tt arXiv:1907.13234}}].

\bibitem{Hidding:2020ytt}
M.~Hidding, {\it {DiffExp, a Mathematica package for computing Feynman
  integrals in terms of one-dimensional series expansions}},  {\em Comput.
  Phys. Commun.} {\bf 269} (2021) 108125,
  [\href{http://arxiv.org/abs/2006.05510}{{\tt arXiv:2006.05510}}].

\bibitem{Armadillo:2022ugh}
T.~Armadillo, R.~Bonciani, S.~Devoto, N.~Rana, and A.~Vicini, {\it {Evaluation
  of Feynman integrals with arbitrary complex masses via series expansions}},
  {\em Comput. Phys. Commun.} {\bf 282} (2023) 108545,
  [\href{http://arxiv.org/abs/2205.03345}{{\tt arXiv:2205.03345}}].

\bibitem{Chen:2023}
T.-Z.~Y. Wen~Chen, Ming-xing~Luo and H.~X. Zhu, {\it {Soft Theorem to Three
  Loops in QCD and ${\cal N}$=4 Super Yang-Mills Theory}},
  \href{http://arxiv.org/abs/2309.03832}{{\tt arXiv:2309.03832}}.

\end{thebibliography}

\providecommand{\href}[2]{#2}\begingroup\raggedright\endgroup

\end{document}